\documentclass[useAMS,usenatbib]{mn2e}

\usepackage{natbib}
\usepackage{graphicx,pifont}
\usepackage{txfonts}
\usepackage{mathrsfs}

\title[A comparison of chemistry and dust cloud formation in ultracool dwarf model atmospheres]
      {A comparison of chemistry and dust cloud formation\\ in ultracool dwarf model atmospheres}
\author[Ch.~Helling et al.]
       {Ch.~Helling,$^{1}$\thanks{E-mail: Christiane.Helling@st-and.ac.uk},  
       A.~Ackerman$^{2}$,  F.~Allard $^{3, 4}$,  M.~Dehn$^{5}$,  P.~Hauschildt$^{5}$,  D.~Homeier$^{6}$,
       \newauthor K.~Lodders$^{7}$, M.~Marley$^{8}$, F. Rietmeijer$^{9}$, T.~Tsuji$^{10}$, P.~Woitke$^{11}$\\
$^{1}$ SUPA, School of Physics and Astronomy, Univ. of St Andrews, North Haugh, St Andrews,  KY16 9SS, UK\\
$^{2}$  NASA Goddard Institute of Space Studies, New York, New York, USA\\
$^{3}$  Centre de Recherche Astrophysique de Lyon, CNRS, UMR5574,
     Universit\'e de Lyon, Ecole Normale Sup\'erieure de Lyon, 
     47 All\'ee d'Italie, F-69634 Lyon, France\\
$^{4}$ Institut d'Astrophysique de Paris, CNRS, UMR 7095,
     98$^{bis}$ Boulevard Arago, F-75014 Paris, France\\
$^{5}$  Hamburger Sternwarte, Gojenbergsweg 112, 21029 Hamburg, Germany\\ 
$^{6}$  Georg-August-Universit\"at G\"ottingen, Institut f\"ur Astrophysik, Friedrich-Hund-Platz, 137077 G\"ottingen, Germany\\
$^{7}$  Planetary Chemistry Laboratory, Department of Earth and Planetary Science,  Washington University, St. Louis, MO, 63130\\
$^{8}$ NASA Ames Research Center, MS 254-3, Moffett Field, CA 94035\\
$^{9}$ Department of Earth and Planetary  Sciences, MSC03-2040, University of New Mexico, Albuquerque, NM 87131-0001,  USA\\
$^{10}$ Institute of Astronomy, The University of Tokyo, 2-21-1 Osawa, Mitaka, Tokyo, 181-0015 Japan\\
$^{11}$ UK Astronomy Technology Center, Royal Observatory, Blackford Hill, Edinburgh EH9 3HJ, Scotland, UK}

\def\tmix{\tau_{\rm mix}} 
\def\etal{${\rm \hspace*{0.8ex}et\hspace*{0.7ex}al.\hspace*{0.5ex}}$}

\begin{document}

\date{Accepted .  Received ; in original form 2008 month day}

\pagerange{\pageref{firstpage}--\pageref{lastpage}} \pubyear{2007}

\maketitle

\label{firstpage}

\begin{abstract}
{ The atmospheres of substellar objects contain clouds of oxides,
iron, silicates, and other refractory condensates.  Water clouds are
expected in the coolest objects.  The opacity of these `dust' clouds
strongly affects both the atmospheric temperature-pressure profile and
the emergent flux.  Thus any attempt to model the spectra of these
atmospheres must incorporate a cloud model.  However the diversity of
cloud models in atmospheric simulations is large and it is not always
clear how the underlying physics of the various models compare.
Likewise the observational consequences of different modelling
approaches can be masked by other model differences, making objective
comparisons challenging.  In order to clarify the current state of the
modelling approaches, this paper compares five different cloud models
in two sets of tests. Test case 1 tests the dust cloud models for a
prescribed L, L--T, and T-dwarf atmospheric (temperature T, pressure p,
convective velocity $v_{\rm conv}$)-structures. Test case 2 compares
complete model atmosphere results for given (effective temperature
T$_{\rm eff}$, surface gravity $\log g$).  All models agree on the
global cloud structure but differ in opacity-relevant details like
grain size, amount of dust, dust and gas-phase composition. These
models can loosely be grouped into {\it high-} and {\it low-altitude
cloud} models whereas the first appear generally redder in
near-infrared colours then the later. Comparisons of synthetic
photometric fluxes translate into an modelling uncertainty in apparent
magnitudes for our L-dwarf (T-dwarf) test case of $0.25 \lesssim
\Delta m \lesssim 0.875$ ($0.1 \lesssim \Delta m \lesssim 1.375$)
taking into account the 2MASS, the UKIRT WFCAM, the Spitzer IRAC, and
VLT VISIR filters with UKIRT WFCAM being the most challenging for the
models. Future developments will need closer links with laboratory
astrophysics, and a consistent treatment of the cloud chemistry and
turbulence.}

\end{abstract}


%

\section{Introduction} 

The atmospheres of L dwarfs are characterised by clouds, formed
principally of silicate, oxide and iron grains, which shape their emergent
spectra. Likewise the atmospheres of the early T dwarfs are
distinguished by the progressive departure of cloud opacity. At even
lower effective temperature, giant-gas planets are again covered in
clouds and other chemical components become important. Any attempt to
derive fundamental properties of these objects from their spectra
hinges on an understanding of the chemistry and physics of clouds.
Yet clouds are inherently difficult to model since they can feedback
into the chemistry and the physics of the entire atmosphere.  Because
of the complexity of this problem a number of independent groups have
taken very different approaches to describe the cloud formation and
the cloud properties of substellar objects as a function of gravity,
effective temperature, and metallicity.  Here we make a first attempt
to compare the quantitative predictions of these various approaches in
order to better understand the models themselves as well as the
uncertainty which remains in application of these models to real
objects.
 

Atmospheric physics classically involves hydrodynamics, radiative and
convective energy transport, and gas phase chemistry. Effects of
magnetic fields are neglected. Ideally, the only free parameters are
the effective temperature T$_{\rm eff}$, the surface gravity $g$,
radius R$_*$ or mass M$_*$, and element abundances $\epsilon_{\rm i}$.
In order to solve such a coupled system of equations in a
computationally reasonable time, assumptions like the hydrostatic
equilibrium, mixing length theory and chemical equilibrium are made.
Inside substellar atmospheres, chemical equilibrium of the gas phase
is justified due to high collision rates between gas-phase
constituents. Irradiation or atmospheric flows may invalidate this
assumption in the upper atmospheric layers.  The validity of
hydrostatic equilibrium and mixing length theory have been studied in
comparison to large eddy simulations for M-type stars (Ludwig et al.\
2002, 2006) and we know from the direct observation of solar system
giant planets at low gravities and effective temperatures that
hydrostatic equilibrium is an appropriate down to very low pressures
($\sim 1\,\rm \mu bar$) in the atmosphere.  The assumption of
hydrostatic equilibrium coupled with the mixing length theory is
computationally extremely efficient and an accurate approximation in
particular if one is aiming at synthetic spectrum calculations. 

The striking difference of substellar atmosphere models compared to
the classical stellar approach is the necessity to model the formation
of clouds and their feedback onto the entire atmosphere.  New physics
needed to be considered and different tribes emerged being inspired by
AGB star dust formation (Helling et al. 2001a; Woitke \& Helling 2003,
2004), by terrestrial cloud formation (Ackerman \& Marley 2001, Cooper
et al. 2003), by measurements for solar system planets (Rossow 1978,
Marley et al. 1999), or driven by practical considerations (Tsuji et
al. 1996 a,b; Allard et al. 2001). The first attempts on cloud
modelling in brown dwarf atmospheres were undertaken by Lunine,
Hubbard \& Marley (1986) and Tsuji et al. (1996 a,b) who suggested the
influence of clouds on the spectral appearance of brown dwarfs. See
also Ackerman \& Marley (2001) for a review and comparison of the
earlier cloud literature.

The overall, phenomenological understanding of cloud formation in
substellar objects has converged to the picture that dust (or
condensates, see Table~\ref{t:distionary}) forms at a certain height
in the atmosphere where it acts as an efficient element sink leaving
behind a depleted gas phase.  The departure of TiO and FeH spectral
lines from the M to the L dwarfs testifies to this process.  The dust
then settles gravitationally taking condensed elements with it.
Convection and atmospheric mixing replenishes the condensing gas,
resulting in a steady state.  The details of this picture, however,
leave room for debate.  It is for example not clear where the dust
precisely starts to form since this depends on the details of the
model assumptions.

Current models generally employ one of two physical approaches to
understand this process. In the first paradigm gas is mixed upward
into higher altitudes. Dust then forms, falls down and meanwhile grows
until it evaporates below the cloud base (Woitke \& Helling 2003).
The second paradigm imagines the opposite limiting case in which the
gas is well mixed from the deep atmosphere only up to a cloud base.
Grains and gas are transported above cloud base by mixing and grains
fall down under the influence of gravity (Ackerman \& Marley 2001,
Allard et al. 2003).  These two branches rely on fundamentally
different model assumptions: (i) the phase-non-equilibrium concept of
kinetic dust formation (Woitke \& Helling 2003, 2004; Helling \&
Woitke 2006, Helling, Woitke \& Thi 2008), and (ii) the
phase-equilibrium concept of thermal stability (Tsuji et al 1996b,
Tsuji 2005, Allard et al. 2001, Ackerman \& Marley 2001, Burrows et
al. 2006; Cooper et al. 2003).  While (ii) represents the end-state
which a dust forming system achieves for $t\rightarrow\infty$, (i)
describes the kinetic process of the formation of cloud particles on
finite timescales limited by mixing and rain-out. The models also
differ in the choice of dust materials which are assumed (i) to form
or (ii) to be present in the atmosphere. Both areas need serious
attention and corresponding material properties should be obtained
either experimentally (see discussion Sect.~\ref{ss:lab}) or from ab
initio calculation (e.g. Jeong et al. 2000, Patzer et al. 2005) which
both are beyond the scope of this paper.  Given these model
conceptions, a number of different model approaches have been
developed to reproduce observed spectra (Tsuji et al 1996a, Tsuji
2005, Allard et al. 2001, Ackerman \& Marley 2001, Burrows et
al. 2006, Cooper et al. 2003, Dehn 2007, Helling et al. 2008a,b) or
providing detailed information on the dust complex itself (Woitke \&
Helling 2003, 2004; Helling \& Woitke 2006, Helling, Woitke \& Thi
2008).

Driven by this diversity in the field, the aim of this paper is to
provide information and to perform comparative studies of models that aim
to describe the dust clouds in substellar atmospheres.  Kleb \& Wood
(2004) demonstrated that such component-based test studies are an
essential part of scientific methods. As the number, $n$, of model
components increases, the interactions amongst them grow as
$n^2/2$. We therefore need to perform verifications on the components
(here: cloud models) but also to use the method of {\it manufactures
solutions}\footnote{{\it Manufactures solutions} are tests with a
known results. In our case, we manufacture the input quantities to a
certain extent and compare the solutions (test case 1,
Sect.\ref{ss:tc1}).} (e.g. Kleb \& Wood 2004) to verify that the
entire system (here: model atmosphere) attains its theoretical
order-of-accuracy properties.  This goes beyond what has been and
could be provided in the literature so far.

\begin{table}
\caption{Definition and units of quantities. The quantities plotted in Figs.\ref{fig:InpStruc}--\ref{fig:ColCol}
are highlighted in {\bf  boldface}.}
\begin{center}
\begin{tabular}{l|p{4.2cm}p{0.3cm}|l}
\hspace*{-0.2cm}{\bf Quantity} & {\bf Definition} & & {\bf Unit}\\
\hline
$T$                      & {\bf local gas temperature} & \hspace*{-0.8cm}(Fig.~\ref{fig:InpStruc},~\ref{fig:P2CloudStruc}) & K\\
$p_{\rm gas}$   & {\bf local gas pressure }      & & dyn\,cm$^{-2}$\\
$v_{\rm conv}$ & {\bf convective velocity  }     & & cm\,s$^{-1}$\\
$\epsilon_{\rm i}$ & gas element abundance  && (i = H, He, ...)\\
$\epsilon_{\rm i}^0$ & deep element abundance  &&\\
$S$                           &supersaturation ratio \\
\hline 
$\rho_{\rm d}/\rho_{\rm gas}$ & {\bf dust to gas mass ratio} & \hspace*{-0.5cm}(Fig.~\ref{fig:RhoDGAmeanTC1}) &\\
$\rho_{\rm d}$                            & dust mass density & & g\,cm$^{-3}$\\
                                                   & $=\sum_{\rm s} n_{\rm s} M_{\rm s}$\,\, for $a=$ const &\\
                                                   & $=\sum_{\rm s} \int_a f(a) M_{\rm s}(a)\,da$ &\\
s                                                  & dust species & \\
                                                   & (e.g. Fe[s], Mg$_2$SiO$_4$[s], $\ldots$) &\\
$n_{\rm s}$                                & number of dust particles of kind $s$ & &  cm$^{-3}$\\
$a$                                             & grain size                      & & cm\\
$M_{\rm s}$                             & mass of dust particle of kind s  & & g\\
\hline
$f(a)$                                         & grain size distribution: & & cm$^{-3}$cm$^{-1}$ \\
& number of dust grains $n$ per grain size $a$ and per gas volume & & \\
$\langle a\rangle$                   & {\bf mean particle size} & \hspace*{-0.5cm}(Fig.~\ref{fig:RhoDGAmeanTC1}) & cm\\
                                                    & $\displaystyle=\frac{\sum_{\rm s} \int_a f_{\rm s}(a) a \,da}{\sum_{\rm s} \int_a f_{\rm s}(a)\,da}$ &\\
                                                    & for fixed $a=a_0$: & \\
&   $f(a) = \delta (a - a_0) \Rightarrow \langle a\rangle=a_0$ &\\
\hline
$V_{\rm s}/V_{\rm tot}$           &{\bf volume fraction of dust kind  s} & \hspace*{-0.5cm}(Fig.~\ref{fig:VsTC1})& 1 \\
$V_{\rm s}$                               & total dust volume of dust kind s & & cm$^3$ \\
$V_{\rm tot}$                             & total dust volume                          & &cm$^3$\\
\hline
$F(\lambda)$  & {\bf surface flux} & \hspace*{-.8cm}(Fig.~\ref{fig:spectotal1800},~\ref{fig:spectotal1000}) & erg\,s$^{-1}$cm$^{-2}$\AA$^{-1}$\\
$\lambda$     &wavelength & & $\mu$m\\
$\log F_{\rm c}$ & {\bf photometric flux} & \hspace*{-.6cm}(Fig.~\ref{fig:spec05-3.0_1800}) &  erg\,s$^{-1}$cm$^{-2}$\AA$^{-1}$\\
                   & flux convolved with filter c & \\
                   & $\displaystyle=\frac{\int_{\lambda_1}^{\lambda_2} F(\lambda)\, \mbox{trans}_{\rm c}(\lambda)\,d\lambda}{\int_{\lambda_1}^{\lambda_2} \mbox{trans}_{\rm c}(\lambda)\,d\lambda}$ & \\
$\log F_{c0}$    & reference photometric flux (Vega)\\
trans$_{\rm c}(\lambda)$ & function of a photometric filter c \\
                 & between $\lambda_1$ and $\lambda_2$ (see Table~\ref{tab:photFlux})&\\
$m_1 - m_2$ & {\bf colour} & \hspace*{-.65cm}(Fig.~\ref{fig:ColCol}) & \\
            & $\!\!\!\displaystyle = 2.5 \big(\log \frac{F_{\rm c2}(\Delta \lambda_2)}{F_{c02}(\Delta \lambda_2)} - \log \frac{F_{\rm c1}(\Delta \lambda_1)}{F_{c01}(\Delta \lambda_1)}\big)$ &\\
$m$              & apparent magnitude\\
                 & $= -2.5 \log\int_{\lambda_1}^{\lambda_2} F(\lambda) \mbox{trans}_{\rm c}\,d\lambda$\\

\end{tabular}
\end{center}
\label{tab:definitions}
\end{table}%

Our paper begins with a summary of the various dust cloud models. We
provide for the first time a comparative presentation of the different
approaches concerning chemistry and dust modelling
(Sect.~\ref{s:cd}). Based on a workshop held in Leiden in October 2006
(\footnote{http://www.lorentzcenter.nl/lc/web/2006/203/info.php3?wsid=203}$^,$\footnote{http://phoenix.hs.uni-hamburg.de/BrownDwarfsToPlanets1/}),
we present test cases where we first separate the components for
chemistry and dust cloud modelling from the complete atmosphere
problem (Sect.~\ref{ss:tc1}). This allows us to judge the
order-of-accuracy properties of the complete models with respect to
chemistry and dust formation which both are essential ingredients for
the solution of the radiative transfer problem.  Section~\ref{ss:tc2}
demonstrates the results for the complete substellar atmosphere
problem, synthetic photometric fluxes, and colours are calculated and
synthetic trust ranges derived from independent models are given.

Comparative studies have been carried out for simulations of radiative
transfer (Pascucci et al. 2004, Iliev et al. 2006), of white dwarfs
(Barstow et al. 2001) or photon dominated region (R\"ollig et
al. 2007).  No comparison study has been presented so far for
substellar atmospheres (brown dwarfs and planets). While we compare a
number of model predictions, including emergent spectra, we refrain
from comparing against spectra of individual substellar objects, since
there are as yet no such objects with independently constrained mass, age, and
metallicity against which spectral models can be compared.


\section{Approaches to chemistry and dust cloud modelling in Brown Dwarf atmospheres }\label{s:cd}

\subsection{Gas phase chemistry}\label{ss:gpalone}

Each of the cloud models to be summarised
(Sect.~\ref{ssec:dustmodels}) assumes local thermodynamic equilibrium
(LTE) when modelling the gas phase chemistry.  In stellar atmospheres,
departures from LTE can arise from interactions of atoms and molecules
with the non-thermal radiation field (Woitke, Kr\"uger \& Sedlmayr
1996) but this effect is negligible in dense substellar atmospheres
(Hauschildt et al. 1997; Schweitzer, Hauschildt \& Baron 2000).  Dust
and gas are assumed to have the same temperature $(T_{\rm dust}=T_{\rm
gas})$.  The gas phase abundances determine the kind and the amount of
dust condensing and are in turn determined by the amount of elements
not bound by the dust.  All codes use equilibrium constants Kp in
their gas-phase treatments (Fegley \& Lodders 1994; Tsuji et
al. 1996b, Tsuji 2005; Allard et al. 2001, 2003; Lodders 2003; Helling
\& Woitke 2006; Helling, Woitke \& Thi 2008). In reality, differences
may arise due to the selection of the input data, which would be
apparent in direct comparisons among the various gas-phase models.

The aim of this paper is to investigate and quantify the differences
arising from different cloud model approaches rather than testing
thermodynamic data sources.  While we do not expect large
uncertainties due to possible difference in thermodynamical gas-phase
data, the results of the gas-phase chemistries used by different
modellers will differ if different sources for element abundances were
used (see discussion in Sect.~\ref{ss:tc1}).

\hspace*{-1cm}
\begin{table*}
\caption{Dust cloud models in substellar atmospheres ($z$ - atmospheric height; $s$ - dust species).\newline
The references are the following: \ding{172} -- Tsuji (2000); \ding{173} -- Tsuji  et al. (1996b); \ding{174}  -- Tsuji (2002, 2005),
\ding{175} -- Allard et al. (2001); \ding{176} -- Allard et al. (2003); \ding{177} -- Ackerman \& Marley (2001), \ding{178} -- Woitke \& Helling (2003), Helling, Woitke \& Thi (2008) }
\begin{tabular}{l|l|l|l|l|lc}
{\bf Author} & \multicolumn{3}{c|}{{\bf Assumptions}}& \multicolumn{3}{c}{\bf Model variants} \\
            &  grain size $a$ & grain composition & supersaturation&  &  & references\\
\hline
Tsuji          & $a=10^{-2}\mu$m & homogeneous & $S=1$ & {\it case B}  & full dusty model   & \ding{172}\\
               &                 &             &       & {\it case C}  & dust cleared model & \ding{173}\\
               &                 &             &       & {\it UCM}     & dust between       & \ding{174}\\
               &                 &             &       &               & $T_{\rm cr}<T<T_{\rm cond}$ \\ [0.2cm]
Allard \& Homeier & $f(a)=a^{-3.5}$ (ISM)   & homogeneous & $S=1$ & {\it dusty} & full dusty model & \ding{175}\\
               &                 &             &       & {\it cond}    & dust cleared model & \ding{175}\\
               & time scale comparison & homogeneous & $S=1.001$ & {\it settl} & & \ding{176}\\
               & ($\nearrow$ Rossow 1978)      &       &               &                    & \\[0.2cm]
Ackerman \& Marley  & log-normal $f(a,z)$ & homogeneous & $S=1$ & $f_{\rm sed}$ & sedimentation& \ding{177}\\
                    & & & & & efficiency \\[0.2cm]
Helling \& Woitke   & $f(a, z)$  & dirty       & $S=S(z,s)$&    &                           & \ding{178}\\
\end{tabular}
\label{tab:modeldetails}
\end{table*}%

\subsection{Dust cloud models}\label{ssec:dustmodels}

In the following we summarise five different cloud models which are
used in substellar atmosphere simulations, and which are involved in
our comparative calculations
(Sect.~\ref{s:results1},~\ref{s:results2}).  While there are many
differences, ultimately all of the models face the same underlying
physical challenges.  We will try to note conceptual similarities
and differences as we describe the models below. 

\noindent
The descriptions include where appropriate:
\begin{itemize}
\item {\it the link between the cloud module and  the atmosphere code},
\item {\it the physical ideas and their representation},
\item {\it the treatment of the cloud chemistry}.
\end{itemize}

\subsubsection{Tsuji model}\label{ss:tsuji}

\paragraph{Dust and gas-phase  treatment in model atmosphere code:}

Tsuji and collaborators apply the methods of non-grey radiative
transfer (in hydrostatic and radiative-convective equilibrium under
LTE) to dusty photospheres with almost no modification, except that
the solid and liquid phases are considered in addition to the gas
phase in chemical equilibrium. In solving chemical equilibrium, the
Tsuji models aim to provide the abundances of ions, atoms, molecules,
and dust grains that contribute to the opacities rather than to derive
a complete solution for all elements.  Thirty-four elements are
considered in charge conservation, 16 elements (H, C, N, O, Na, Mg,
Al, Si, P, S, Cl, K, Ca, Ti, V, Fe) in molecular formation, and 8
elements (Mg, Al, Si, Fe, K, Ca, Ti, V) in dust formation.  The
chemical equilibrium computation includes 83 molecules and is based on
a previous examination of about 500 molecular species (as for details,
including the thermochemical data, see Tsuji 1973).  Dust grains
composed of Fe, Si, Mg, and Al in form of metallic iron, enstatite
(MgSiO$_3$), and corundum (Al$_2$O$_3$) are considered as sources of
dust opacity. The abundances are solved as being in phase equilibrium
with gaseous species. No other dust species composed of Fe, Si, Mg, or
Al are considered for simplicity. Also, abundances of some gaseous
species important as sources of gaseous opacity suffer large reduction
by the dust formation, and such effects are approximated by perovskite
(CaTiO$_3$) for Ti, melilite (Ca$_2$MgSi$_2$O$_7$) for Ca, VO(cr) for
V, and K$_2$S(cr)/KCl(cr) for K, since gaseous TiO, CaH, VO, and K are
important sources of gas opacity.

\paragraph{Cloud model:}\label{tsuji:cloudmodel}
The Tsuji models assume that dust forms in the photosphere as soon as
the thermodynamical condition for condensation is met, i.e. the
supersaturation $S=1$. Then the layers cooler than the condensation
temperature ($T_{\rm cond}$) are assumed to be filled with dust grains
({\it case B}) which act as element sink and opacity source.  Another
extreme case is that the dust grains all precipitate as soon as they
are formed and the atmosphere is thus clear of dust ({\it case C}),
hence the dust acts as element sink but not as opacity source.
Finally an intermediate case (the ``Unified'' or UCM case) in which
grains condense at $T_{\rm cond}$, but precipitate at a slightly lower
temperature termed the critical temperature, $T_{\rm cr}$ is also
considered. In this case the dust cloud appears in a restricted region
of $ T_{\rm cr} < T < T_{\rm cond}$.

While $T_{\rm cond}$ is well defined by thermal stability, $T_{\rm
cr}$ is left as a free parameter to be estimated empirically.  If
$T_{\rm cr}$ is equal to $T_{\rm cond}$, all the dust grains will
precipitate as soon as they are formed ({\it case C}). On the other
hand, if $T_{\rm cr}$ is as low as the surface temperature, all the
dust grains formed will survive in the fully dusty photosphere ({\it
case B}). If $T_{\rm cr}$ differs only slightly from $T_{\rm cond}$,
the dust cloud will be quite thin while the dust cloud will be rather
thick if $T_{\rm cr}$ is much lower than $T_{\rm cond}$. Thus $T_{\rm
cr}$ is essentially a measure of the thickness of the dust cloud and
thus has a significant effect on the infrared colours (not unlike
$f_{\rm sed}$ in the Ackerman \& Marley (2001) model). As a free
parameter $T_{\rm cr}$ (along with $T_{\rm eff}$) can be inferred from
the observed infrared colours. For this purpose, reasonably accurate
values of $T_{\rm eff}$ can be inferred from the luminosities based on
the observed parallaxes and bolometric fluxes (e.g. Golimowski et
al. 2004; Vrba et al. 2004). But it appeared that the infrared colours
differ significantly even for the same $T_{\rm eff}$ (e.g.  Marley et
al. 2005; Tsuji 2005) and this fact implies that $T_{\rm cr}$ also
differs for the same $T_{\rm eff}$. For example, four cool dwarfs from
spectral type L6.5 to T3.5, whose infrared spectra are quite
different, appear to have almost the same empirical $T_{\rm eff}$ at
about 1400$\pm$100\,K.  Such very different spectra of almost the same
$T_{\rm eff}$ could be explained reasonably well by assuming different
values of $T_{\rm cr}$, i.e. different thickness of the dust cloud
(see Fig. 10 of Tsuji 2005).

In the Tsuji models all grains have radius $a = 0.01\,\mu$m.  
In the limit of such small sizes the dust opacity is independent
of the particle size for a fixed mass of dust.

\begin{table*}
\caption{A brief dictionary for multiple meanings and phrases (a, b, $\ldots$)  used by different authors\newline (T - Tsuji; AH -- Allard \& Homeier; MAL - Marley, Ackerman \& Lodders; HW - Helling  \& Woitke; R -- Rietmeijer). }
\begin{center}
\begin{tabular}{p{2.5cm}|cl|p{12cm}}
dust         & a) & (HW)  & general term for small solid particles, grains, liquid droplets\\
             & b) & (MAL) & condensate\\
condensation & a) & (T, AH)   & dust formation by conversion of vapour to solid (or liquid) particles\\[0.2cm]
nucleation   & a) & (MAL, HW) & seed particle formation\\
growth       & a) & (HW)      & formation of condensate species by chemical surface reaction on an existing surface\\
evaporation  & a) & (HW)      & reverse growth process ($\tau_{\rm evap}$ - evaporation time scale)\\ [0.2cm]
drift        &  a) & (HW)     & relative motion between gas and dust\\
             &  b) & (MAL)    & gravitational settling\\ 
             &  c) & (MAL)    & rain, rain-out\\
             &  d) & (AH, MAL)& sedimentation: the falling of cloud particles under the influence of gravity ($\tau_{\rm sed}$ - sedimentation time scale) \\
             &  e) & (T, MAL) & precipitation: formation of cloud particle for which $\tau_{\rm evap}\gg\tau_{\rm sed}$ (from Rossow 1978)\\[0.2cm]
homogeneous \,\,\,\,nucleation  &  a) & (HW) & seed formation by addition of the same monomer species forming clusters of increasing sizes until they achieve solid state character\\
              &  b) &  & formation of first condensate that will grow  to increasingly larger clusters \\
heterogeneous & a) &   & seed formation by addition of different monomer species\\
nucleation    & b) & (R) & formation of condensed species onto an existing seed ($\nearrow$ growth)\\
primary condensate & a) & (MAL) &  a condensate forming from the gas by gas-gas reactions\\
secondary condensate & a) & (MAL) & a condensate forming by gas-solid reaction with previously existing solid or liquid phase\\[0.2cm]
coagulation   & a) &  & formation of one particle from two colliding cloud particles\\
coalescence   & a) & (AH) & coagulation caused by size dependent sedimentation velocities of cloud particles of different sizes\\
element conservation & & & exchangeable used with mass balance since no elements should be created inside the atmosphere\\
supersaturation & a) & (HW, T, AH)  & $S$, the ratio of the monomer particle pressure to the saturation vapour pressure\\
                &    &              & (see Helling, Woitke \& Thi 2006, Appendix for discussion) \\
                & b) & (MAL) & $S-1$, the vapour pressure in excess of saturation divided by the saturation vapour pressure
\end{tabular}
\end{center}
\label{t:distionary}
\end{table*}%

\subsubsection{Allard \& Homeier model}\label{ss:lyon}

\paragraph{Dust and gas-phase  treatment in model atmosphere code:}
The Allard \& Homeier models solve for chemical equilibrium in the gas
phase by minimisation of the functional errors, where the functions
are the elemental and charge conservation, Saha equation, and
mass-action law for each of 40 elements, with up to 5 ionisation
levels per atom, and for some 600 molecules and nearly as many
condensate using thermochemical data from many sources including a
compilation of the JANAF tables (Chase et al. 1986; for details see
Allard et al. 2001 and Allard \& Hauschildt 1995).

Allard et al. (2001) modelled the limiting effects of cloud formation
({\it dusty}, {\it cond}) on the spectral properties of late M and L
to T brown dwarfs by treating dust in chemical equilibrium with the
gas phase.  For the grains construction and opacities in the {\it
dusty} and the {\it cond} models, an interstellar size distribution of
spherical and chemical homogeneous grains
is assumed. A slight supersaturation ($1.001$) is assumed in the {\it
settl} models (Allard et al. 2003, 2007).

The grain sizes are calculated from the comparisons between time-scales for
mixing due to convective
 overshooting as prescribe by (Ludwig et al. 2002) and condensation and
 gravitational settling
 according to Rossow (1978).
 The thermal structure of the model atmosphere is solved on a fixed
optical depth grid at $1.2\mu$m assuming no dust opacity contribution.
The {\it dusty} models accounted for dust opacity while
the {\it cond} models did not (Allard et al. 2001).  The {\it settl}
models involve a detailed cloud model (Allard et al. 2003, 2007) which is
solved for the thermal structure of the atmosphere to find the grain
size and abundances distributions as function of depth. In order to account
for the gas cooling effects as it is propelled by convective
turbulence from the top of the convection zone towards the top of the
atmosphere, the cloud model is solved by depleting gas phase abundances
 layer-by-layer from the
 innermost (assumed of solar composition) to the outermost layer.


The resulting stratified elemental abundances and number densities of
species are then used in the radiative transfer solver applying the
Mie equation and  complex refractive index for 
calculating the dust opacities (Ferguson et al. 2005).  Models are
converged removing thereby any possible cloud opacity inconsistencies
between thermal structure and radiative transfer.

\paragraph{Cloud model:}\label{HomAll:cloudmodel}
For the {\it settl} model atmosphere (Allard et al. 2003, 2007) in each
layer, the condensation, sedimentation and coalescence timescales
(Rossow 1978; see Table~\ref{t:distionary}) are compared to the mixing
timescale prescribed by Ludwig et al. (2002) as follows:

\begin{description}
\item[a)]  the equilibrium size between mixing and
  sedimentation is calculated and  the growth time scale (condensation and
  coalescence) is computed for that size;
\item[b)] the mixing time scale is then compared to the growth time
  scale:
  \begin{itemize}
  \item if  growth is faster the condensates are found to be depleted, 
  and  the fraction of  condensates is recomputed so as to obtain a 
  growth time scale equal to the mixing time scale;
  \item when mixing is faster, the growth is limited by the 
  replenishment with fresh condensable material from deeper layers,
  and, while the condensate fraction is stable,  a 
  mean size is recomputed corresponding to an equilibrium between mixing and 
  condensation.
  \end{itemize}
\end{description}

Given the new cloud description the elemental abundances are then 
readjusted which produces a new equilibrium
condensate fraction.  These steps are  repeated  until the
condensate fraction no longer changes.  This is a time consuming
process which however guaranties that the chemistry reflects the
cooling path of the gas.

Another essential input to the model is the description of the mixing
timescale. Within the lower classically convective unstable atmosphere
layers a mixing velocity is readily obtained from the results of
mixing length theory, which is implemented in \texttt{PHOENIX}\ in the
formulation of Mihalas (1978). Since the principal cloud formation
region is located well above the Schwarzschild boundary, one is
confronted with the task of extrapolating the velocities over several
pressure scale heights.  For a phenomenological description of this
velocity field this group draws on the results of the hydrodynamical
simulations of late M dwarfs by Ludwig et al. (2002), which show in
general an exponential decline of mass transfer by overshooting with
decreasing pressure, after an initial transition zone.  Further
simulations by Ludwig (2003) indicate a steepening of this decline
with surface gravity. The mass exchange frequency following these
simulations is parameterised in analogy to the Helling \& Woitke
-model as $\log \tau_\mathrm{mix}(z) = \log \tau_\mathrm{MLT} + \beta
(\log p_0 - \log p(z)$, where the base value of the mixing time scale
(where $p=p_0$) within the convectively unstable layer is given by its
mixing length theory value $\tau_\mathrm{MLT} = \alpha
H_p/v_\mathrm{conv}$ and $\alpha=2.0$ the mixing length parameter
(Ludwig et al. 2002).  The slope can be derived from Ludwig (2003) as
$\beta = 2 \sqrt{g_5} $, where $g_5$ is the surface gravity in units
of $10^5\mathrm{cm\,s}^{-2}$.  Since the calibration of this relation
involved an extrapolation of the M dwarf simulations to lower
T$_{\rm eff}$, the models allow for adjusting the slope $\beta$ by a
factor of up to 3, adopting a factor of 1 for the 1800K and 1400K test
cases, and 2 for the 600K and 1000K cases.

\subsubsection{Marley, Ackerman \& Lodders model}\label{ss:ames}


\paragraph{Dust treatment in model atmosphere code:}
The Marley, Ackerman \& Lodders - modelling treats the upward
convective mixing of a gas, its condensation, and the sedimentation of
condensate through the atmosphere of an ultracool dwarf.  The
composition and cloud structure at each point in a trial atmosphere
model is computed, based on the existing profile and then this
information is used to iterate towards the next trial profile.  The
chemistry at each pressure/temperature point is interpolated within a
table of atomic and molecular abundances computed for chemical
equilibrium (Freedman, Marley \& Lodders 2008).  The cloud is computed
by applying the Ackerman \& Marley (2001) cloud model.

\paragraph{Gas-phase treatment:} Abundances of gas species are
calculated with the CONDOR code (Lodders \& Fegley 1993; Fegley \&
Lodders 1994; Lodders 2003) which calculates chemical equilibrium
compositions by considering the dual constraints of mass balance and
chemical equilibrium. Input data required for the code are
thermodynamic properties of the gas-phase species and compounds (e.g.,
equilibrium constants), appropriate elemental abundance tables for the
system, temperature and total pressure. The equilibrium constants
used in the CONDOR code are computed from the Gibbs free energy
($\Delta\,G$) data, which are directly proportional to the logarithm
of the equilibrium constant ($\ln K_{\rm p}$) as $\Delta G = -RT \ln K_{\rm p}$
($T$ -- gas temperature, $R$ -- gas constant). The code considers $\sim
2000$ gas species (including ions) and $\sim 1700$ solids and liquids
for compounds of all naturally occurring elements\footnote{All
elements up to Bi (number 83) excepting Tc and Pm but adding Th and U,
for a total of 83.}.

\paragraph{Condensate treatment:}
For application to substellar atmospheres, the CONDOR code treats
condensate formation by removing primary condensates (i.e.,
condensates that form from condensing gases) from the gas into cloud
layers (Lodders 2004, Lodders \& Fegley 2006; see
Table~\ref{t:distionary}).  An important consequence of this approach is
that secondary condensates arising from gas-solid reactions as would
be predicted by pure equilibrium are excluded because the primary
condensates are assumed to settle into clouds and are thus no longer
available for reaction with the cooler gas above the clouds. For
example the computation assumes that iron grains (a primary
condensate) do not react with $\rm H_2S$ gas to form FeS at lower
temperatures ($\sim 700\,\rm K$) where the secondary FeS would form if
Fe metal were still present.  Instead the $\rm H_2S$ remains in the
gas phase as is observed in the deep atmosphere of Jupiter (Niemann et
al. 1998) where $\rm H_2S$ is only removed into $\rm NH_4SH$ clouds
below $\sim 200\,\rm K$ (Fegley \& Lodders 1994; Visscher et
al. 2006). Likewise a detection of $\rm H_2S$ in a cool T dwarf would
confirm the inhibition of secondary condensation.  Marley et
al. (2002) argue that the far red colours of T dwarfs can only be
reproduced if secondary condensation of alkali-bearing phases is
indeed inhibited.

Both the cloud model of the settling of primary condensates and the
chemical equilibrium model assume that at a given temperature below
the condensation temperature the gas phase abundances of the elements
sequestered by condensation are set by the respective vapour pressure
of the primary condensate. With this common assumption the cloud and
chemical computations are fully self-consistent.

\paragraph{Cloud model:}\label{Marely:cloudmodel}
The cloud model (Ackerman \& Marley 2001) parameterises the
efficiency of sedimentation of cloud particles relative to turbulent
mixing through a scaling factor, $f_{\rm sed}$ (Eq. 4 in Ackerman
\& Marley 2001).  Large values of $f_{\rm sed}$ correspond to rapid
particle growth and large mean particle sizes.  In this case
sedimentation is efficient, which leads to physically and optically
thin clouds.  When $f_{\rm sed}$ is small particles are assumed to
grow more slowly and the amount of condensed matter in the atmospheric
is larger and clouds thicker.  In this sense small $f_{\rm sed}$
is somewhat comparable to the Tsuji models with a large difference
between $T_{\rm cond}$ and $T_{\rm crit}$ while large $f_{\rm sed}$ is similar to the 
opposite case.  Unlike the Tsuji models Ackerman \& Marley (2001) compute
a particle size profile for each condensate in each model atmosphere.

For a fixed atmospheric profile,
$f_{\rm sed}$ and a description of the width of the particle size
distribution, the Ackerman \& Marley (2001) model uniquely predicts
the variation in mean particle size and particle number density
through the atmosphere.  Thus families of models, i.e. set of models
with varying sets of parameters T$_{\rm eff}$, $\log g$, f$_{\rm
sed}$, each with a unique $f_{\rm sed}$, can be produced.
No attempt is made to model microphysical processes of dust growth 
and coagulation.  Instead it is assumed that the micro-physical processes
acting within the cloud are able to produce the particle sizes
implied by any specified value of $f_{\rm sed}$.

In terrestrial rain clouds the particle size distribution is often
double-peaked (Ackerman et al. 2001), with small particles that grow
from condensation of the vapour co-existing with larger drops that
have grown by by collisions between particles.  In the Ackerman\&
Marley (2001) approach a single, broad log-normal particle size
distribution is intended to capture the likely existence of such a
double-peaked size distribution.  They employ a width $\sigma = 2$ for
all cases, although this can in principle be varied.

Like the Allard \& Homeier - model, the Marley, Ackerman \& Lodders cloud - model
use the mixing length theory to compute gas velocities in the
convection zone and must employ some other description to specify
velocities above the radiative-convective boundary. They describe
mixing in radiatively stable layers by specifying an eddy diffusion
coefficient, $K_{\rm eddy} = H^2/\tau_{\rm mix}$, where H and
$\tau_{\rm mix}$ are the scale height and mixing time.  Experience
with the radiative stratospheres in the solar system (e.g., Atreya et
al. 1991; Bishop et al. 1995; Moses et al. 2004)
shows that typical values of $K_{\rm eddy}$ in these atmospheres lie
in the range of $10^4$ to $10^7\,$cm$^2$s.  Observations of ammonia
and CO in the atmosphere of the T7.5 dwarf Gliese 570D imply $K_{\rm
eddy}\sim10^6\,$cm$^2$s (Saumon et al. 2006; Geballe et al. 2008).
Comparisons of the mid-infrared colours of L dwarfs to models that
include chemical mixing (Leggett et al. 2006), suggest $K_{\rm
eddy}\sim10^4\,$cm$^2$ s$^{-1}$.  The Marley, Ackerman \& Lodders - models
reported here set $K_{\rm eddy}\ge10^5$\,cm$^2$ s$^{-1}$ at all points
in the atmosphere with a smooth transition from the convective zone to
this value.

As with variation of cloud thickness to match variation in $J-K$ at
fixed $T_{\rm eff}$ in the Tsuji models, changes in $f_{\rm sed}$
produce atmosphere models with a range of near-infrared spectra and
colours. Burgasser et al.\ (2007) and Stephens et al.\ (2008) have
shown that the spectra of bluer-than-average L dwarfs can be fit by
models employing large $f_{\rm sed}$ while redder-than-average L
dwarfs seem to require small $f_{\rm sed}$.  The Marley, Ackerman \&
Lodders reproduce spectra across the L to T transition by
employing models with progressively larger $f_{\rm sed}$ with later
spectral type (Cushing et al. 2008; Stephens et al. 2008; Leggett et
al. 2008).

\subsubsection{Helling \& Woitke model}\label{ss:wh}

\paragraph{Dust treatment for a model atmosphere code:}
The Helling \& Woitke approach is fundamentally different from the
previous models in two important ways.  First, this model follows the
trajectory of an ensemble of dust grains downwards from the top of the
atmosphere instead of upwards from the bottom.  This approach is based
on the phenomenological analogy to thunderstorm where large air masses
are advected upwards before raindrops do form. Dust clouds in
substellar objects are considered stationary, i.e. uncondensed gas is
mixed upward from which dust particles continuously form, settle
gravitationally, and evaporate. In this stationary situation, the
downward directed element transfer via precipitating dust grains is
balanced by an upward mixing from the deep interior by
convective  and overshoot-motions (Helling et al. 2001b, Helling
2003). 
The second major difference from the other approach is that the
Helling \& Woitke approach kinetically describes the cloud particle
formation as phase-transition process by modelling seed formation,
grain growth/evaporation, sedimentation in phase-non-equilibrium,
element depletion, and their interactions.  Dust moment equations
describing these processes are derived from rate equations and are
solved as a function of height $z$ for a given $(T, p, v_{\rm conv})$
atmosphere structure. The equations are integrated inward.

\paragraph{Gas-phase treatment:}
The composition of the gas phase is calculated assuming chemical
equilibrium for 14 elements (H, He, C, N, O, Si, Mg, Al, Fe, S, Na, K,
Ti, Ca) and 158 molecules with equilibrium constants fitted to the
thermodynamical molecular data of the electronic version of the JANAF
tables (Chase et al. 1986). The equilibrium constant for TiC are from
Gauger et al. (see Helling et al. 2000), for CaH from Tsuji (1973),
and FeH from Burrows (priv. com.). Solar elemental abundances are
assumed at the lower boundary of the model atmosphere, assuming a
well-mixed gas-phase solar composition, and first ionisation states of
the elements are calculated.  Element conservation equations are
auxiliary conditions which take into account the loss of elements
in the gas phase by nucleation, growth, and drift and the gain by
evaporation (Woitke \& Helling 2004).

\paragraph{Condensate treatment:}
The condensates considered during the solution of the Helling \& Woitke dust model
equations are treated in full phase-non equilibrium.  The
supersaturation ratio $S$ is calculated from the 
gas phase composition in chemical equilibrium.

\paragraph{Cloud model:}\label{HeWo:cloudmodel}
The dust formation starts with the formation of seed particles
(nucleation). The nucleation rate is calculated for homogeneous
(TiO$_2$)$_N$-clusters applying the modified classical nucleation
theory (Gail et al. 1984; see Eq. 34 in Helling \& Woitke 2006). The
calculation of the nucleation rate relies on quantum
mechanical calculations for the formation of TiO$_2$-seeds by a
step-wise addition of TiO$_2$ molecules (Jeong et al. 2000).  The
nucleation rate determines the number of dust particles. These seeds
grow to macroscopic sizes by gas-solid surface reactions. Because many
compounds can be thermally stable almost simultaneously in substellar
atmospheres, the simultaneous growth of 12 solids TiO$_2$[s], SiO[s],
SiO$_2$[s], Fe[s], FeO[s], Fe$_2$O$_3$[s], FeS[s], MgO[s],
MgSiO$_3$[s], Mg$_2$SiO$_4$[s], Al$_2$O$_3$[s] and CaTiO$_3$[s] by 60
surface reactions onto TiO$_2$-seed particles is modelled (Helling,
Woitke \& Thi 2008).  These {\it dirty grains} are modelled to be
composed of a homogeneous mix of numerous islands of the different,
pure condensates (Helling \& Woitke 2006). Drift transports existing
particles into region where they might continue to grow before they
evaporate in the deeper, warmer atmosphere. While reactions on an
existing grain surface proceed if the gas is supersaturated $(S>1)$
with respect to this particular reaction (Helling \& Woitke 2006), the
seed formation can only take place when the gas is highly
supersaturated $(S\gg1)$. If the gas is under-saturated $(S<1)$ the
solid will evaporate.

The majority of dust grains that build up the cloud layer are found in
a subsonic gas for which Knudsen numbers are small (see Woitke \&
Helling 2003). The respective kinetic description is solved in form of
conservation equations which allows a simultaneous treatment of
nucleation, growth, evaporation, drift, and element replenishment.
The dust formation is modelled by applying conservation equations of
dust moments $L_{\rm j}=\int V^{\rm j/3} f(V) dV$ with $f(V)$ the
grain size distribution function. Nucleation, growth/evaporation and
gravitational settling are source terms of these equations (Woitke \&
Helling 2003, Helling \& Woitke 2006).  The solution of the dust
moment equations and element conservations determines quantities like
grain sizes, grain material composition, total grain volume, remaining
gas-phase element abundances. The element replenishment is treated by
introducing a parameterised mixing time scale $\tmix(z)$.  Ludwig\etal
(2002, 2006) show that, generally speaking, the convectively excited
hydrodynamical motions -- and thereby the mixing -- decay
exponentially with increasing height above the convectively unstable
zone resulting in an exponential decrease of the mass exchange
frequency in the radiative zone from which $\tmix(z)$ is derived
as $\log \tmix(z) = \log\tmix^{\rm min}(z) + \beta\cdot\{\,0, \log
p_o - \log p(z)\,\}$ with $p_0$ the pressure at the upper edge of the
convective unstable zone, $\tmix^{\rm min}(z)=\alpha/H_{\rm p} v_{\rm
conv}$ ($\alpha=2.0$) the minimum value of the mixing time-scale
occurring in the convectively unstable region and $\beta=\Delta\log
f_{\rm exchange}/\Delta\log p\approx 2.2$.

\subsubsection{Dehn \& Hauschildt + Helling \&  Woitke model}\label{dhhw}

\paragraph{Dust treatment in model atmosphere:}
The dust cloud model of Helling \& Woitke (Sect. \ref{ss:wh}) has been
adopted as module in the static {\sc Phoenix} model atmosphere code
(Dehn 2007; Helling et al. 2008 a,b). the dust module receives the
$(T(z), p(z), v_{\rm conv}(z))$ structure from {\sc Phoenix} and
provides the dust number density, the solid's volume fractions, the
mean grain size, and the remaining element abundances in the gas phase
for each atmospheric layer.  Effective medium and Mie theory are then
used to calculate the dust opacity in addition to the usual gas-phase
opacity calculations. The temperature structure is found iteratively
by a modified Uns\"old-Lucy correction algorithm.  The adjusted
atmosphere structure, including the solution of mixing-length theory
to find $v_{\rm conv}$, is an input for the dust module in the next
iteration.  Compared to the classical {\sc Phoenix} solution, the
computing time has increases by a considerable factor of since for
each temperature iteration the dust module is called. The dust module
itself iterates to solve the dust moment equation by fulfilling the
element conservation auxiliary condition.

\paragraph{Gas-phase treatment, condensate treatment, cloud model:} The
gas-phase composition is calculated assuming chemical equilibrium as
described in Sect.~\ref{ss:lyon}. The cloud model is a reduced version
of Sect.~\ref{ss:wh} in order to keep the computation time effortable:
The simultaneous growth of 7 solids TiO$_2$[s], SiO$_2$[s], Fe[s],
MgO[s], Mg$_2$SiO$_3$[s], Mg$_2$SiO$_4$[s], Al$_2$O$_3$[s] onto the
TiO$_2$-seed particles is considered. Only 32 surface reaction are
taken into account (Dehn 2007).

  \begin{figure}
   \centering
   \includegraphics[width=8.5cm]{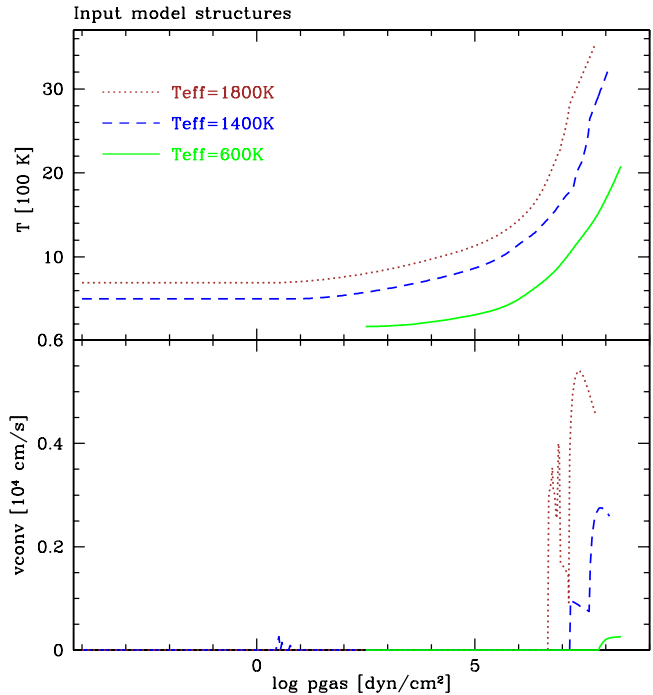}
      \caption{Input model structure $(T, p, v_{\rm conv})$ for
              T$_{\rm eff}=$1800, 1400, 600 K with $\log\,$g=5.0 and
              solar element abundances.}
         \label{fig:InpStruc}
   \end{figure}

\section{Test cases}\label{s:tcs}

We explored the characteristics and capabilities of the cloud models summarised in
Sect.~\ref{s:cd} with two sets of test calculations. The
{\bf test case 1} is the component-based test study (Kleb \& Wood
2004), and it is designed to compare the dust cloud models alone by
separating them from hydrodynamics and radiative transfer treatments
(including opacity calculations).  {\bf Test case 2} utilises the method
of manufactured solutions (Kleb \& Wood 2004), and compares the
results of completely iterated substellar model atmosphere
simulations.  Table~\ref{tab:definitions} contains the definitions of
the quantities discussed in the following.

\subsection{Test case 1: local quantities given} \label{ss:tc1}

Each dust cloud model is calculated for a prescribed set of ($T,p_{\rm
gas},\epsilon^0_{\rm i}, v_{\rm conv}$) or ($T,p_{\rm gas},
\epsilon^0_{\rm i}, F_{\rm conv}$) profiles with $T$ the local gas
temperature, $p_{\rm gas}$ the local gas pressure, $v_{\rm conv}$ and
$F_{\rm conv}$ being the convective velocity and the convective flux,
respectively (Fig.~\ref{fig:InpStruc}).  The deep, well-mixed element
abundances $\epsilon^{\rm 0}_{\rm i}$ (i=H, Si, Mg, Ti, $\ldots$) in
the inner atmosphere have been chosen as solar according to Grevesse,
Noels \& Sauval (1992).

\medskip
\noindent
{\sc Remarks on the solar element abundances:} The solar element abundances
published by Anders \& Grevesse (1989), Grevesse, Noels \& Sauval
(1992), and Grevesse \& Sauval (1998) are undergoing considerable
revisions but an agreement on the most correct value according to
present knowledge has not yet been reached.  The oxygen element
abundance, as the most prominent example, has been revised downward to
$\epsilon^0_O=8.66\pm0.05$ by Asplund et al~(2004) and to only
$\epsilon^0_O=8.76\pm0.07$ by Caffau et al~(2008), both based on 3D
hydrodynamical simulations of the solar photosphere in combination
with non-LTE line transfer (see discussion in Caffau et
al.~2008). However, Ayres et al. (2006) suggest $\epsilon^0_O=8.84$
from their measured solar CO lines. Additionally, the downward
revision of the oxygen abundances greatly increases the difference
between the internal sound speed predicted by solar models and the
sound speed inferred from helioseismology (Christensen-Dalsgaard et
al. 2008).  The determination of solar element abundances is a
fundamental problem for atmosphere physics and chemistry, and the
final amount of dust formed in a cloud will depend on the element
abundance values. However, the test of its implications goes beyond
the scope of this paper. Therefore, {\bf test case 1}
(Sect.~\ref{s:results1}) applies the Grevesse, Noels \& Sauval (1992).
The element abundances used for {\bf test case 2}
(Sect.~\ref{s:results2}) are listed in Table~\ref{tab:codes} for each
of the atmosphere codes.

\medskip
\noindent
{\sc Remarks on the mixing time-scale $\tau_{\rm mix}$:} A mixing
time-scale $\tau_{\rm mix}$ enters all cloud models except the
Tsuji-model. Each  model (Marley, Ackerman \& Lodders,
Allard \& Homeier, Helling \& Woitke) does interpret, and hence,  uses
$\tau_{\rm mix}$ in different ways. In the Marley, Ackerman \&
Lodders-model, the sedimentation efficiency is parameterised relative
to a mixing time-scale through a scaling factor $f_{\rm sed}$, the
Allard \& Homeier-model includes a $\tau_{\rm mix}$ in a time-scale
comparison to determine local mean grain sizes, and a $\tau_{\rm mix}$
influences the rate of seed formation, the growth and the settling
process since it enters a set of conservation equations in the Helling
\& Woitke-model. Hence, we refrain from directly comparing the mixing
time-scales of the different cloud models, since a comparison of $\tau_{\rm
mix}$ would be rather misleading regarding the cloud properties
derived by each of the modeller groups.  

\medskip
\noindent
We have chosen to compare the models for the following stellar
parameter which can be considered as examples for the L\,--, L\,--\,T,
and T--dwarf atmospheres:
\begin{tabbing}
 L\,--\,dwarf   \hspace*{0.5cm}         \= T$_{\rm eff}$= 1800K   \=  (provided by M. Dehn)\\ 
  L\,--\,T dwarf  \> T$_{\rm eff}$= 1400K   \> (provided by M. Dehn) \\
 T\,--\,dwarf  \hspace*{0.5cm} \>  T$_{\rm eff}$= 600K \hspace*{0.5cm} \>(provided by M. Marley)\\*[-0.5cm] 
\end{tabbing}

All models have $\log g=5.0$. The given $(T, p)$ and $(v_{\rm conv},
p)$ structures are shown in Fig.~\ref{fig:InpStruc}. The T$_{\rm
eff}$= 600K model is considerably less extended in $\log$\,p than the
hotter models. Its convective velocity is very small. Therefore, a
much less efficient convective overshooting is anticipated, and 
(for those models which assume gaseous transport) 
a less efficient element replenishment of the upper atmospheric
layers.

\subsection{Results: test case 1}\label{s:results1}

We compare four essential results of our dust cloud models which are
 needed for the opacity calculations in a complete atmosphere model:
\begin{itemize}
\item {\it dust content} 
\item {\it mean particle size}
\item {\it dust material composition}
\item {\it gas-phase composition}
\end{itemize}

  \begin{figure*}
   \centering
 \hspace*{-0.5cm}  \includegraphics[width=9cm]{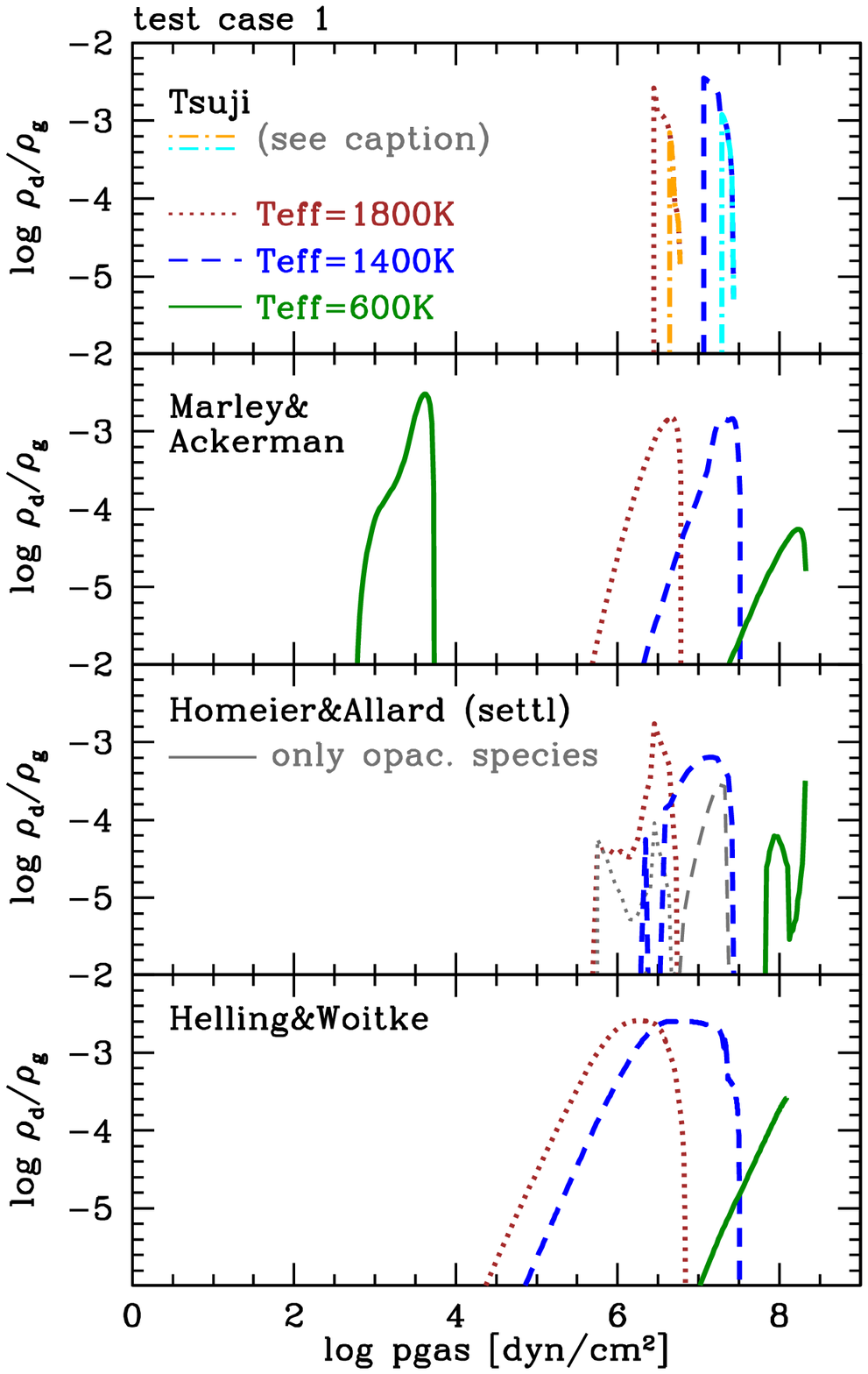}\hspace*{0.2cm}
   \includegraphics[width=9cm]{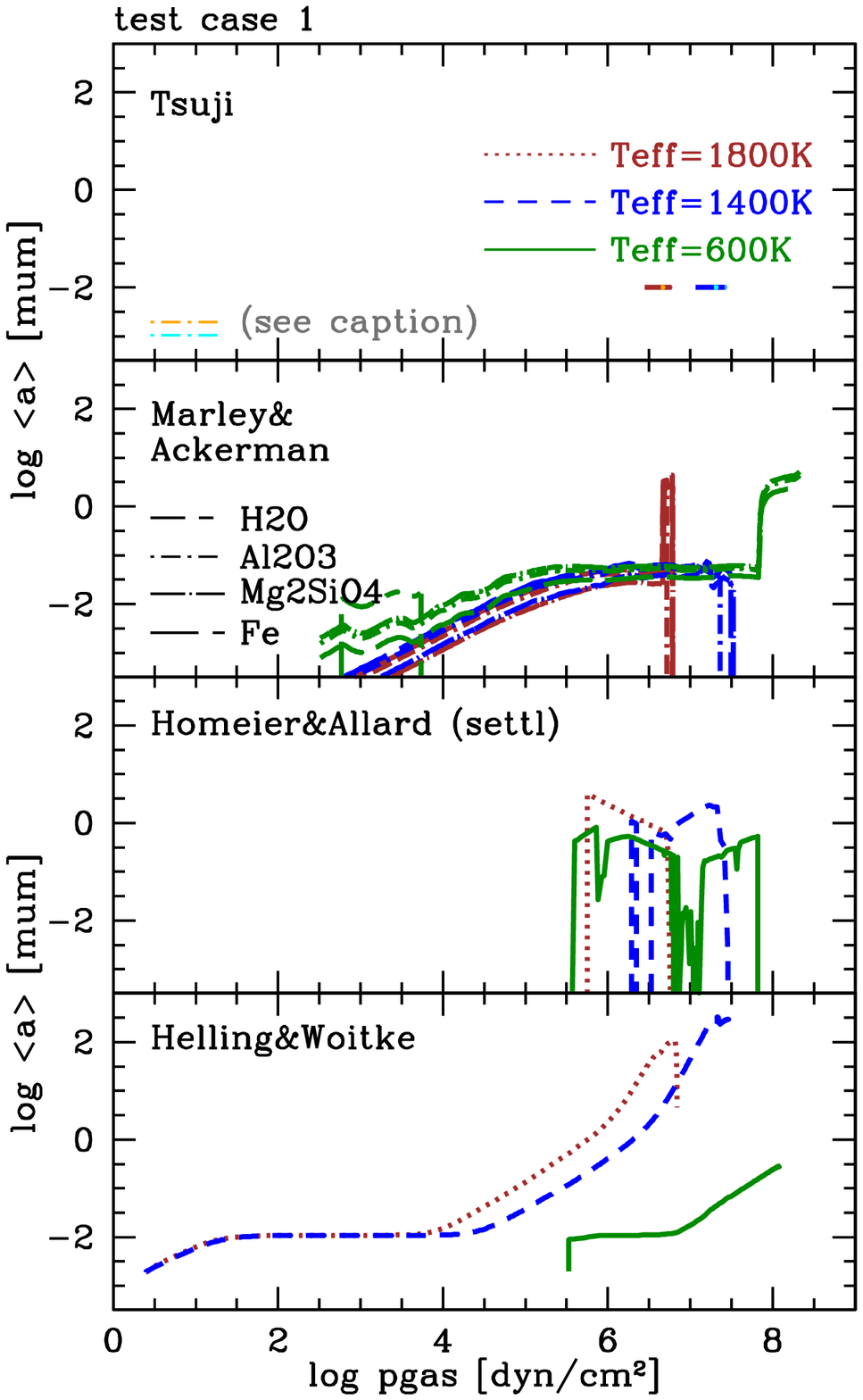}
      \caption{Test case 1 results for prescribed atmospheric
structures for T$_{\rm eff}$=1800, 1400, 600K, all $\log$\,g = 5.0 and
solar element abundance. {\bf Left:} Dust-gas-ratio $\rho_{\rm
d}/\rho_{\rm g}$ {\bf Right:} Mean particle size $\langle a\rangle$
[$\mu$m]\newline {\bf Note:} For Tsuji two cases of $T_{\rm cr}$ are
plotted for each $T_{\rm eff}$: $T_{\rm cr}=1900$K -- light colours
(orange/light blue), $T_{\rm cr}=1700$K -- dark colours
(red/blue). For Marley, Ackerman \& Lodders $\rho_{\rm d}/\rho_{\rm g}
= (\rho_{\rm Al2O3} + \rho_{\rm Fe} + \rho_{\rm Mg2SiO4} + \rho_{\rm
H2O})/\rho_{\rm g}$ ($f_{\rm sed}=2$). The homogeneous H$_2$O-,
Al$_2$O$_3$-, Mg$_2$SiO$_4$-, and Fe-particle have different sizes
(different line styles). For Allard \& Homeier $\rho_{\rm d}/\rho_{\rm
g}$ includes all species contributing to the depletion of the gas
phase. The gray lines show their values which enters the radiative
transfer calculation. For the Allard \& Homeier T$_{\rm
eff}=600$K-model, the opacity-species-only-$\log \rho_{\rm
d}/\rho_{\rm g}$ values fall below the axis range depicted. For
Helling \& Woitke, the code has difficulties calculating clouds for
T$_{\rm eff}=600$K in the inner atmosphere.}
         \label{fig:RhoDGAmeanTC1}
   \end{figure*}

  \begin{figure*}
   \includegraphics[width=5.8cm]{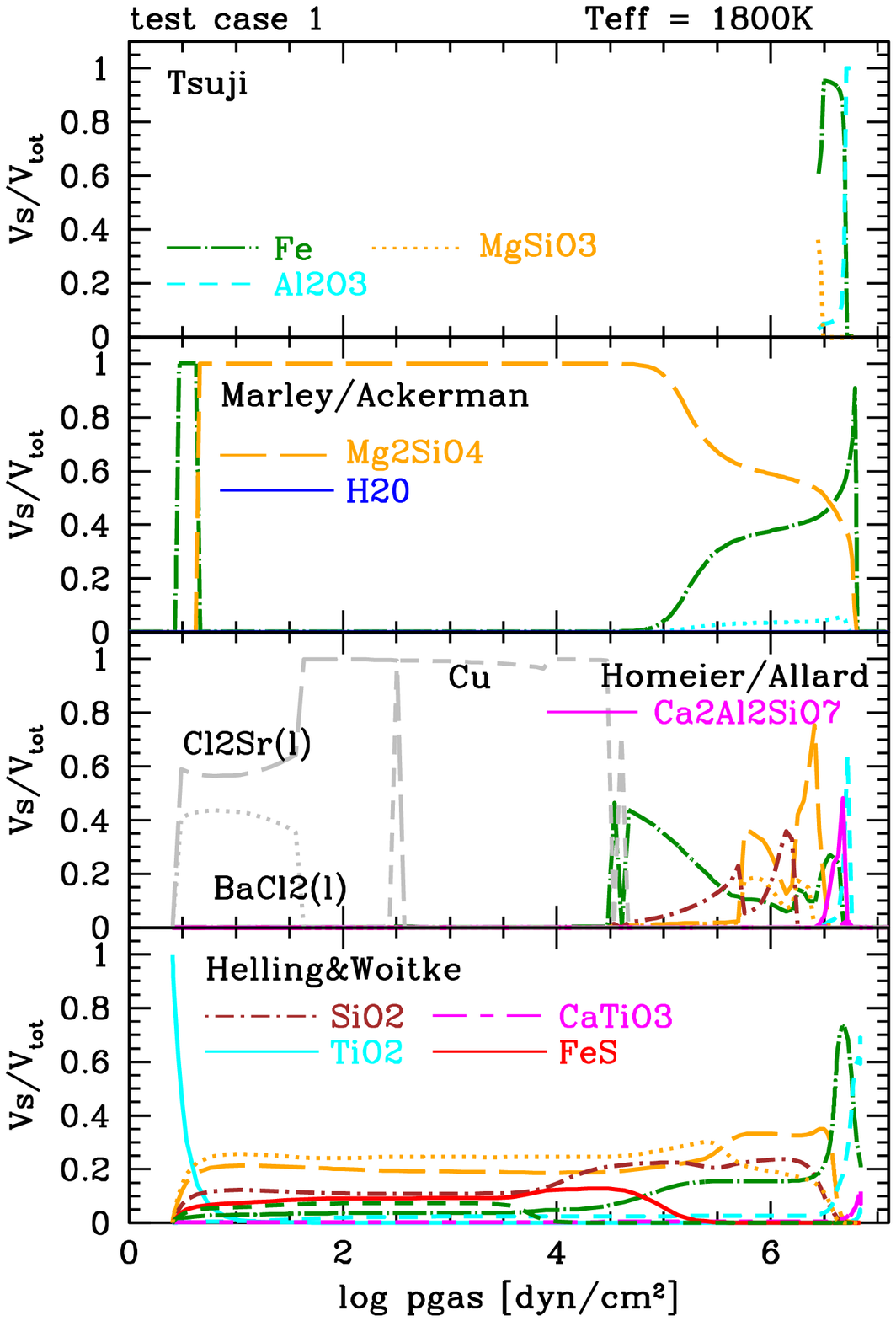}
   \includegraphics[width=5.8cm]{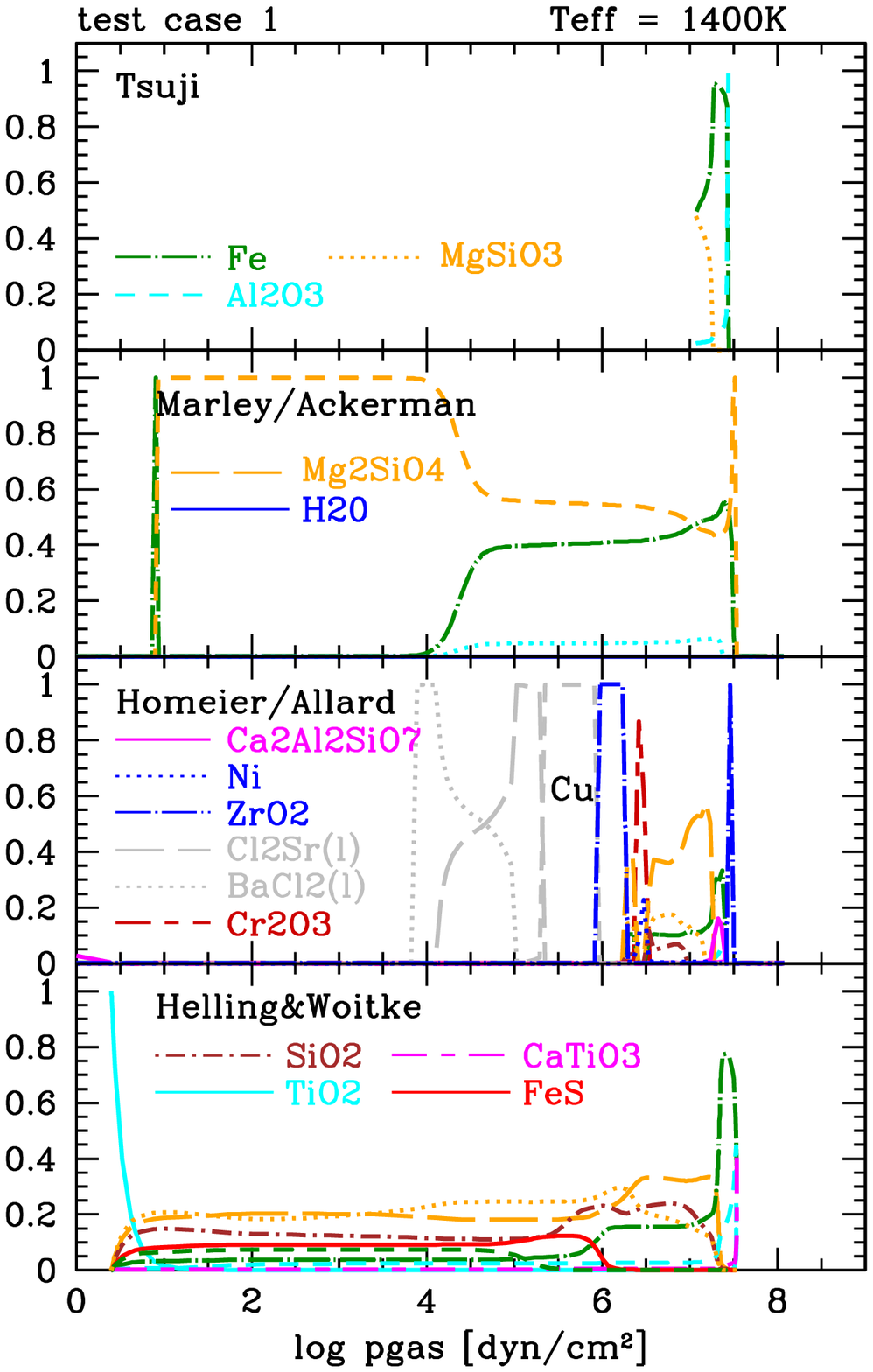}
   \includegraphics[width=5.8cm]{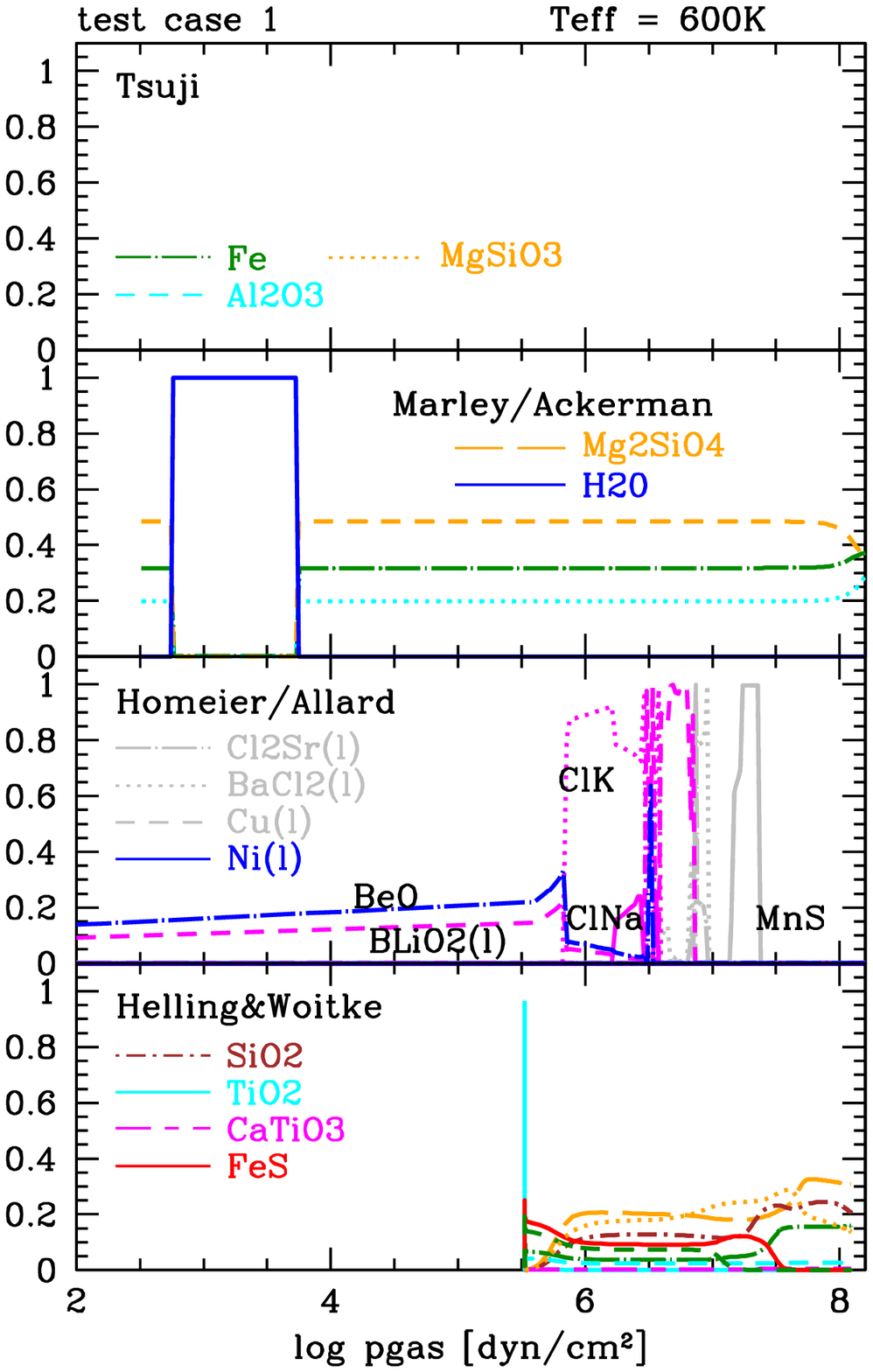}
      \caption{Cloud material composition in volume fractions of total
 dust volume $V_{\rm s}/V_{\rm tot}$ for prescribed model atmospheres
 of T$_{\rm eff}$=1800, 1400, 600K, $\log$\,g = 5.0 (Tsuji: T$_{\rm
 cr}=1700$K,Marley, Ackerman \& Lodders: $f_{\rm sed}=2$) resulting
 from different dust cloud models.\newline {\bf Note:} For Allard \&
 Homeier $V_{\rm Al2O3}/V_{\rm tot} =(V_{\rm Al2O3-c1}+ V_{\rm
 Al2O3-c2}+ V_{\rm Al2O3-c3})/V_{\rm tot}$ and $V_{\rm Fe}/V_{\rm tot}
 = (V_{\rm Fe-c}+ V_{\rm Fe-c1})/V_{\rm tot}$ with c, c1, c2, c3 being
 different crystal structures treated in their equilibrium ansatz. For
 Tsuji no T$_{\rm eff}=600$K model is available.}
   \label{fig:VsTC1}
   \end{figure*}

\subsubsection{Dust content in the atmosphere}\label{ss:rhorhog}

We measure the dust content in the atmosphere by the dust-to-gas mass
ratio $\rho_{\rm dust}/\rho_{\rm gas}$ (for definition see
Table~\ref{tab:definitions}; Fig.~\ref{fig:RhoDGAmeanTC1} left).
However, the phase-equilibrium models allow two interpretation of this
quantity, namely, the amount of dust acting as opacity source and the
amount of dust acting as element sink (compare column 6 in
Table~\ref{tab:codes}).  Figure~\ref{fig:RhoDGAmeanTC1} depicts
$\rho_{\rm dust}/\rho_{\rm gas}$ for dust opacity sources, and
demonstrate for the Allard \& Homeier-model that the difference to
$\rho_{\rm dust}/\rho_{\rm gas}$ for dust element sinks can be
significant.

The dust-to-gas mass ratio shows where most of the dust is located in
the cloud layers, and the extension of the cloud layer(s) differs for
all models. The innermost cloud layers generally contain the maximum
amount of dust, except in the Tsuji results where the clouds extension
varies with T$_{\rm cr}$.  All models have the same location of the inner
boundary of the cloud ({\it cloud base}), since it is determined mainly
by thermal stability.

 The maximum $\rho_{\rm dust}/\rho_{\rm gas}$ is of about the same
order of magnitude for all models but the exact values differ (see
Table~\ref{tab:rhodg}).  Note that the amount of dust entering the
radiative transfer calculation is usually smaller than the amount of
dust causing the gas phase depletion in phase-equilibrium models
(compare Allard \& Homeier-model: gray lines in
Fig.~\ref{fig:RhoDGAmeanTC1}, left).  The maximum $\rho_{\rm
dust}/\rho_{\rm gas}$ value is reached at different atmospheric
altitudes in the different models and it retains its value over
different atmospheric extension.  The Tsuji-models suggest the highest
amount of dust in the atmospheres for $T_{\rm cr}=1700$K.  The Allard
\& Homeier-models suggest the lowest amount of dust. The $T_{\rm
eff}=600$K test case is challenging for all models: No data could be
provided from the Tsuji-model, the Allard \& Homeier $\rho_{\rm
dust}/\rho_{\rm gas}$ have a local minimum, and the Helling \&
Woitke-model reaches the shallowest depth.  The Marley, Ackerman \&
Lodders $T_{\rm eff}=600$K-model (solid green line in
Fig.~\ref{fig:RhoDGAmeanTC1}) shows two well-separated cloud layers: a
water layer high up in the atmosphere at $\approx 10$ dyn/cm$^2$ and a
silicate layer between $10^7\,\ldots\,10^8$ dyn/cm$^2$ (compare
Sect.~\ref{ss:Vs}). All other models produce only the silicate layer.

\begin{table}
\caption{Maximum dust-to-gas ratios  $\log\,(\rho_{\rm dust}/\rho_{\rm gas})_{\rm max}$,  the maximum mean grain sizes $\log\,\langle a \rangle$  [$\mu$m] and its value in the upper cloud layers @ $10^3$ dyn/cm$^2$ in different dust models for given ($T,p_{\rm gas}, v_{\rm conv}$)  profiles.}
\begin{tabular}{p{1.5cm}p{0.35cm}|p{0.5cm}p{1.6cm}p{1.1cm}p{1.1cm}}
                        & &  &   Marley, & Allard\&  &Helling\& \\
& \hspace*{-0.3cm} T$_{\rm eff}$  & Tsuji  &Ackerman & Homeier &  Woitke \\
                       &         & & \& Lodders & & \\
\hline
dust content & \hspace*{-0.3cm}1800K   & $-3.6$                              & $-3.2$  & $-3.2$ & $-3.4$ \\
$\log \rho_{\rm dust}/\rho_{\rm gas}$ & \hspace*{-0.3cm}1400K       & $-3.6$ & $-3.2$   & $-2.8$ & $-3.4$ \\
& \hspace*{-0.3cm}600K                                                 &   & $-5.8$ {\tiny (silicates)} &$-4.5$  & $-4.4$\\
                      & &   & $-3.5$ {\tiny  (H$_2$O)} & &\\
\hline
\hline
grain size & \hspace*{-0.3cm}1800K & $-2$ & \hspace*{-0.2cm} $-1.4\,\ldots\,-1.6$ & 0 & $-2$ \\
$\log\,\langle a \rangle$\,\,\, @ & \hspace*{-0.3cm}1400K & $-2$ & \hspace*{-0.2cm} $-1.4\,\ldots\,-1.6$ & 0 &  $-2$ \\
 $10^3$ dyn/cm$^2$                                 & \hspace*{-0.3cm}600K   &  &\hspace*{-0.2cm} $-1.6\,\ldots\,-2$ & 0 & 0\\
\hline
maximum &  \hspace*{-0.3cm}1800K & $-2$  &  $+0.5$ & $+0.7$ & $+2.0$\\
    grain size &\hspace*{-0.3cm}1400K &  $-2$  &   1 & $+0.5$ &  $+2.5$\\
     $\log\,\langle a \rangle_{\rm max}$    &  \hspace*{-0.3cm}600K   &  $-2$  &  $+0.5$ & 1 &  $-0.5$
\end{tabular}
\label{tab:rhodg}
\end{table}

\subsubsection{Mean particles sizes in the cloud layer}

Figure~\ref{fig:RhoDGAmeanTC1} (right) shows the results for the means
grain sizes $\langle a \rangle$ (definition see
Table~\ref{tab:definitions}) calculated for given ($T,p, v_{\rm
conv}$) profiles. The mean grain sizes are different amongst all
models which reflects the different model assumptions
made. Also, the grain size distribution function $f(a)$ used
to determine $\langle a \rangle$ is different in each of the dust
cloud models (see Table~\ref{tab:modeldetails}).

A common feature for all models is that small mean particle sizes
$\langle a \rangle \lesssim 10^{-2}\mu$m populate the upper cloud
regions, except in the Allard \& Homeier-model.  This small grain size
in the upper cloud layers are associated with very small dust-to-gas
ratios of $\rho_{\rm dust}/\rho_{\rm gas} < 10^{-6}$.  Particle sizes
increase inward and reach a certain maximum size (Marley, Ackerman \&
Lodders; Helling \& Woitke), or are constant by assumption in the
entire cloud (Tsuji), or they reflect a complicated time-scale
competition (Allard \& Homeier). The grains of different kind s have
different distributions $f_{\rm s}(a,z)$ in Marley, Ackerman \&
Lodders-model (i.e. Fe[s]-grains, H$_2$O[s] grains -- different line
styles in Fig.~\ref{fig:RhoDGAmeanTC1}, $3^{\rm rd}$ panel). Grains of
different but homogeneous composition have the same size distribution
in the Allard \& Homeier and in the Tsuji-models at a particular
height in the atmosphere.  The dirty grains (i.e. a mixture of
Fe[s]-SiO[s]-Mg$_2$SiO$_4$[s] etc.)  in the Helling \& Woitke-model
are characterised by one mean grain size distribution $f(a,z)$ at a
particular height z in the atmosphere. The transition from $\langle a
\rangle_{\rm min} $ to $\langle a \rangle_{\rm max} $ across the cloud
layer appears smoothly in the Helling \& Woitke models.  The Marley,
Ackerman \& Lodders- models reach their $\langle a \rangle_{\rm max} $
abruptly at about the same latitude. Also the Allard \& Homeier models
show a sudden rise in mean grain size but at a different atmospheric
height. Another direct consequence of the different dust cloud
modelling is that the values of $\langle a \rangle_{\rm max} $ can
differ by a factor of 100 (see also Table~\ref{tab:rhodg}).

\subsubsection{Dust material composition}\label{ss:Vs}

The chemical composition of the cloud particles shows the largest
variation between the different models. Figure~\ref{fig:VsTC1} shows
the material composition of the test models T$_{\rm eff}=1800, 1400,
600$\,K in volume fractions of the total dust volume, $V_{\rm
s}/V_{\rm tot}$ (definition $\nearrow$ Table~\ref{tab:definitions}),
for the different cloud models.  We only consider dust species which
are important for the gas and dust opacity in the radiative transfer
calculations in Sect.~\ref{ss:tc2} ($\nearrow$ Table~\ref{tab:codes}).
The calculation of the dust composition varies widely in the different
models and splits into two classes: dust particles of homogeneous
composition assuming equilibrium condensation (Tsuji; Marley, Ackerman
\& Lodders; Allard \& Homeier) and dirty dust particles of
heterogeneous composition according to the kinetic treatment of growth
and evaporation (Helling \& Woitke). The chemical heterogeneity of the
whole dust complex in the equilibrium models is reached by considering
ensembles of pure Fe[s]-grains, Al$_2$O$_3$[s]-grains etc.

The Tsuji and the Marley, Ackerman \& Ackerman-models allow for the 3
and 5 major condensates, respectively, as dust opacity sources. The
dominating low-temperature condensate is Mg$_2$SiO$_3$[s] in the
Tsuji-model and Mg$_2$SiO$_4$[s] in the Marley, Ackerman \&
Lodders-model. The Fe[s] fraction increases with decreasing T$_{\rm
eff}$ in both models at intermediate temperatures.  Also in the
Helling \& Woitke-model Mg$_2$SiO$_3$[s] and Mg$_2$SiO$_4$[s] are the
dominant low temperature condensates in addition to SiO$_2$[s] and
FeS[s]. Note that SiO$_2$[s] is never predicted by the equilibrium
models.  The Fe[s] content increases inward until it reaches, like in
the other models, a pronounced maximum near the inner cloud edge. The
dust at the cloud base is made of Al$_2$O$_3$[s] with possible
impurities of CaTiO$_3$[s] (T$_{\rm eff}$=1800, 1400K) in the Helling
\& Woitke-model. The most refractory cloud condensate layer in Marley,
Ackerman \& Lodders-models is composed of corundum or Ca-Aluminates
which serve as element sinks, hence, not depicted in
Fig.~\ref{fig:VsTC1}.  The Allard \& Homeier-models only partially
agree with these results.
The Marley, Ackerman \& Lodders-model is the only model which includes
H$_2$O[s] as possible condensate which allows for a second, detached
cloud layer above the already discussed oxide--silicate cloud layer (compare
Fig.~\ref{fig:RhoDGAmeanTC1}).

Given the great diversity in grain composition with different model assumptions,
we must conclude that the chemical composition of the cloud particles
in substellar atmospheres is still uncertain.

 \begin{figure*}
 \center
\includegraphics[width=15cm]{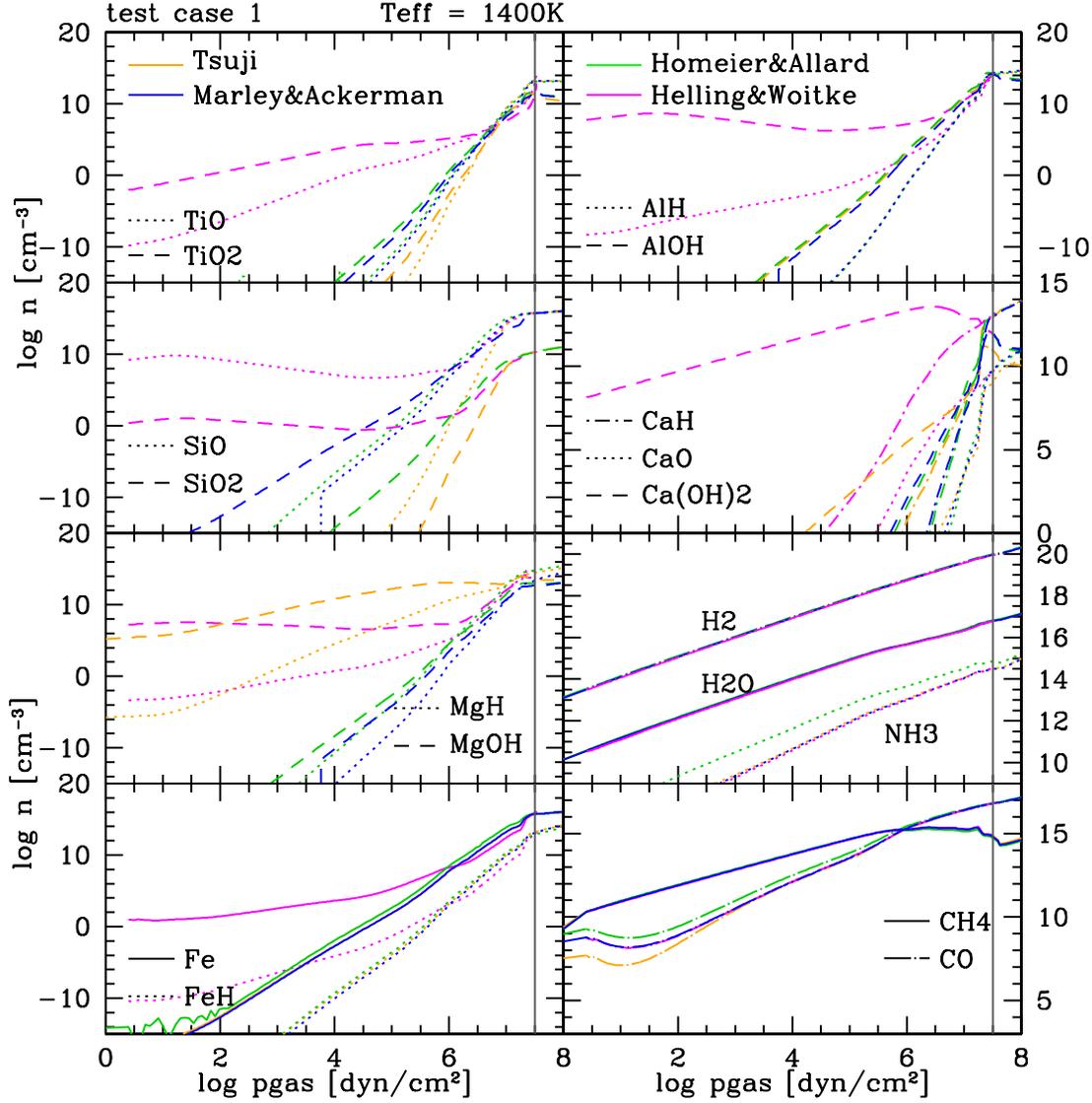}
      \caption{Gas phase composition in a cloud forming atmosphere
      with T$_{\rm eff}$=1400K, $\log\,$g=5.0 (Tsuji: T$_{\rm
      cr}=1700$K, Marley, Ackerman \& Lodders: $f_{\rm sed}=2$) as resulting
      from different cloud model approaches. The vertical thin black
      line indicates the pressure at the cloud base where $\rho_{\rm
      d}/\rho_{\rm gas}\rightarrow 0$ ($\nearrow$
      Fig.~\ref{fig:RhoDGAmeanTC1}, left). }
         \label{fig:GasPhase1400}
   \end{figure*}

\subsubsection{Gas-phase chemistry results}\label{sss:gaph}
Figure~\ref{fig:GasPhase1400} shows the number densities $n$
[cm$^{-3}$] for a selected number of gas phase Si-, Mg-, Al-, Ca-, and
Ti-bearing molecules. We additionally plot the most important
H-molecules and the most important C-bearing molecules (7$^{\rm th}$
and 8$^{\rm th}$ panel). All models assume the gas-phase to be in
chemical equilibrium ($\nearrow$ Sect.~\ref{ss:gpalone}). All models
used the same well-mixed element abundances $\epsilon^0_{\rm i}$ at
the inner boundary of the cloud mode ($\nearrow$
Sect.~\ref{ss:tc1}). Hence, different gas-phase number densities
produced by the models are a consequence of the different treatment of
dust formation which leads to different remaining element abundances
$\epsilon_{\rm i}$ in the gas phase.  The comparison of the remaining
gas phase (after cloud formation) is needed to understand possible
spectral trends in the later test cases of the complete (sub-)stellar
atmosphere model (Sect.~\ref{ss:tc2}).

\paragraph{H$_2$, H$_2$O, NH$_3$:}
The first test for differences in the chemical equilibrium gas-phase
composition considers H$_2$. Because of the continuous community
interest we include H$_2$O, and NH$_3$ for its increasing spectral
importance with decreasing T$_{\rm eff}$ in the substellar regime.
The H$_2$ and H$_2$O abundances are almost identical for all
models. Figure~\ref{fig:GasPhase1400} shows agreement also for the
NH$_3$ abundances except for the Allard \& Homeier-model which
predicts an overabundance of NH$_3$ compared to the other models.

\paragraph{CO, CH$_4$:}
These molecules are only little effected by dust formation, since
carbon solids are not considered in the models under investigation.
Hence, they are a good test for the general agreement of the gas-phase
composition with respect to element abundances and material
constants. However, the consumption of oxygen by the silicates and
oxides does also affect the amount of gas-phase CO, hence indirectly also
CH$_4$, due to oxygen depletion. We observe that all models predict CH$_4$ to be
the major C-bearing molecule above a certain height in the atmosphere
below which CO takes over.  Beside this general agreement amongst the
models, the CO number densities differ above the cloud layer,
most likely resulting from different equilibrium constants for CO and CH$_4$.

\paragraph{TiO, TiO$_2$:}
TiO$_2$ is more abundant than TiO in all models though the relative
difference varies amongst the models. The models do not
agree on the values of the TiO and TiO$_2$ abundances. The
Helling \& Woitke model suggests the highest abundances for both
molecules, the Tsuji-model suggests the lowest abundances. 

\paragraph{SiO, SiO$_2$:}
SiO is more abundant than SiO$_2$ in all models though the relative
difference varies widely amongst the models. All models agree well for
$p>10^7$dyn cm$^{-2}$ which coincides with the pressure-level of the
maximum dust content in this model (compare
Fig.~\ref{fig:RhoDGAmeanTC1}, left). The Tsuji-model again suggest the
lowest molecular abundances, and the Helling \& Woitke model suggest
the highest abundances at lower pressures.

\paragraph{MgH, MgOH:}
MgOH is more abundant than MgH in all models and the number densities
agree well in the inner atmosphere for $p>10^7$dyn cm$^{-2}$. The
molecular abundances of MgH and MgOH fall into two groups with respect
to the upper atmosphere: The Tsuji- and Helling \& Woitke-models
suggest a high number density. The Allard \& Homeier- and Marley,
Ackerman \& Lodders-model predict the lowest abundances.

\paragraph{Fe, FeH; AlH, AlOH:}
All models suggest that atomic Fe is more abundant than molecular FeH,
and AlOH than AlH throughout the atmosphere and agree very well in the
inner atmosphere $p>10^7$dyn cm$^{-2}$. All phase-equilibrium models
yield very good agreements for all four molecules, and the Helling \&
Woitke suggests the highest Fe, FeH, AlH, and AlOH gas-phase
abundances in the outer atmosphere layers.

\paragraph{CaH, CaO, Ca(OH)$_2$:}
All models suggest that Ca(OH)$_2$ is more abundant than CaH for
$p<10^6$dyn\,cm$^{-2}$.  CaO has generally a much lower abundance than
these two molecules. The Helling \& Woitke-model produces exceptionally
high abundances of the Ca-molecules as result of the limited number of
Ca-bearing solids ($\nearrow$ Sect.~\ref{ss:wh}). The Tsuji-model
suggest the next highest abundances for these molecules.

\paragraph{General:} 
The general trend is that the phase-equilibrium models (Tsuji, Allard
\& Homeier, Marley, Ackerman \& Lodders) produce lower gas-phase
abundances of molecules containing dust-forming elements then the
kinetic model (Helling \& Woitke) in the upper atmosphere.  However,
the molecules that are not affected by the chemistry of dust formation
(like CO, CH$_4$, H$_2$O) have very similar abundances in the
different models. However, differences for these molecules are
indicative of the different oxygen-consumption caused by the
differences in the dust cloud models, and of possible differences in
the material quantities ($\nearrow$ Sect.~\ref{ss:gpalone}). Those
molecules containing rare element (like Al, Ti, Ca) are predicted with
very similar abundances in all phase-equilibrium models. Remaining
deviations for these molecules are likely due to a missing solid as
element sink ($\nearrow$ Ca Fig.~\ref{fig:GasPhase1400}). The
strongest deviations amongst these models occurs for molecules
containing very abundant elements (Si, Mg). Since a large fraction of
the Mg- and Si-bearing molecules contributes to the dust formation,
the differences in the dust models are imprinted in the remaining
gas-phase abundances the strongest.  All models agree on the gas-phase
composition below the cloud base (vertical black line,
Fig.~\ref{fig:RhoDGAmeanTC1}, left).

\subsection{Test case 2: global quantities given}\label{ss:tc2}

Two sets of stellar parameters were prescribed,
\begin{tabbing}
 T\,--\,dwarf  \hspace*{0.5cm} \=  T$_{\rm eff}$=1000K, \= $\log\,g=5.0$\\
 L\,--\, dwarf           \> T$_{\rm eff}$=1800K, \> $\log\,g=5.0$,
\end{tabbing}
for which complete model atmospheres were calculated including the
solution of the radiative transfer. All models assume hydrostatic
equilibrium, gas-phase chemical equilibrium, and use mixing length
theory for treating the convective energy transport. The dust cloud
models are those described in Sect.~\ref{s:cd}, and all cloud approaches assume spherical symmetric cloud particles.  Table~\ref{tab:codes}
summarised further details on the atmosphere codes. The cloud modules
are the model atmosphere component which is most different amongst the
codes under consideration in this paper.
 
\subsection{Results test case 2}\label{s:results2}
We compare results for complete stellar atmosphere simulations regarding the
\begin{itemize}
\item {\it atmosphere structure and cloud profile} 
\item {\it spectral energy distribution}
\item {\it photometric fluxes and colours}
\end{itemize}

\begin{table*}
\center
\caption{Summary of the model atmosphere codes used in Sect.~\ref{ss:tc2} (s -- solids only, sl -- liquids and solids)}
\begin{tabular}{cccccccl}
{\bf authors} & {\bf code} & {\bf element}    & {\bf number of} & {\bf number of}       & {\bf number of}    \\
              & {\bf name} & {\bf abundances} & {\bf elements}  & {\bf gas-phase spec.} & {\bf dust species} \\
\hline
{\bf Tsuji}   &           &  Anderse \& Grevesse (1989) & 34 & 83 & 3 as opacity source   & s\\
              &           &  Allende Prieto et al.(2002)&    &      & 10 as element sinks & s\\[0.3cm]
{\bf Allard \& Homeier}  & {\sc Settl-}& Grevesse, Noels \& Sauval (1992)        & 84 & 680 & 43 as opacity source & sl\\
                        & {\sc Phoenix}& Asplund, Grevesse \& Sauval (2005)&    &    &     169 as element sinks   & sl\\[0.3cm] 
{\bf Marley, Ackerman} &              & Lodders (2003)                    & 83 & $\sim 2200$  & 5 as opacity source & s\\
{\bf \& Lodders}                        & &      &    &    & $\sim 1700$ as element sinks & sl\\[0.3cm]
{\bf Dehn \& Hauschildt}   & {\sc Drift-}  &  Grevesse, Noels \& Sauval (1992) & 40 & 338 & 7 as opacity source & s  \\
{\bf + Helling \& Woitke} & {\sc Phoenix} &   &    &     & 7 as element sinks & s\\
\end{tabular}
\label{tab:codes}
\end{table*}

\subsubsection{Atmosphere structure and cloud profiles}\label{ss:atmprof}

\begin{figure*}
\hspace*{-0.5cm}   \includegraphics[width=9cm]{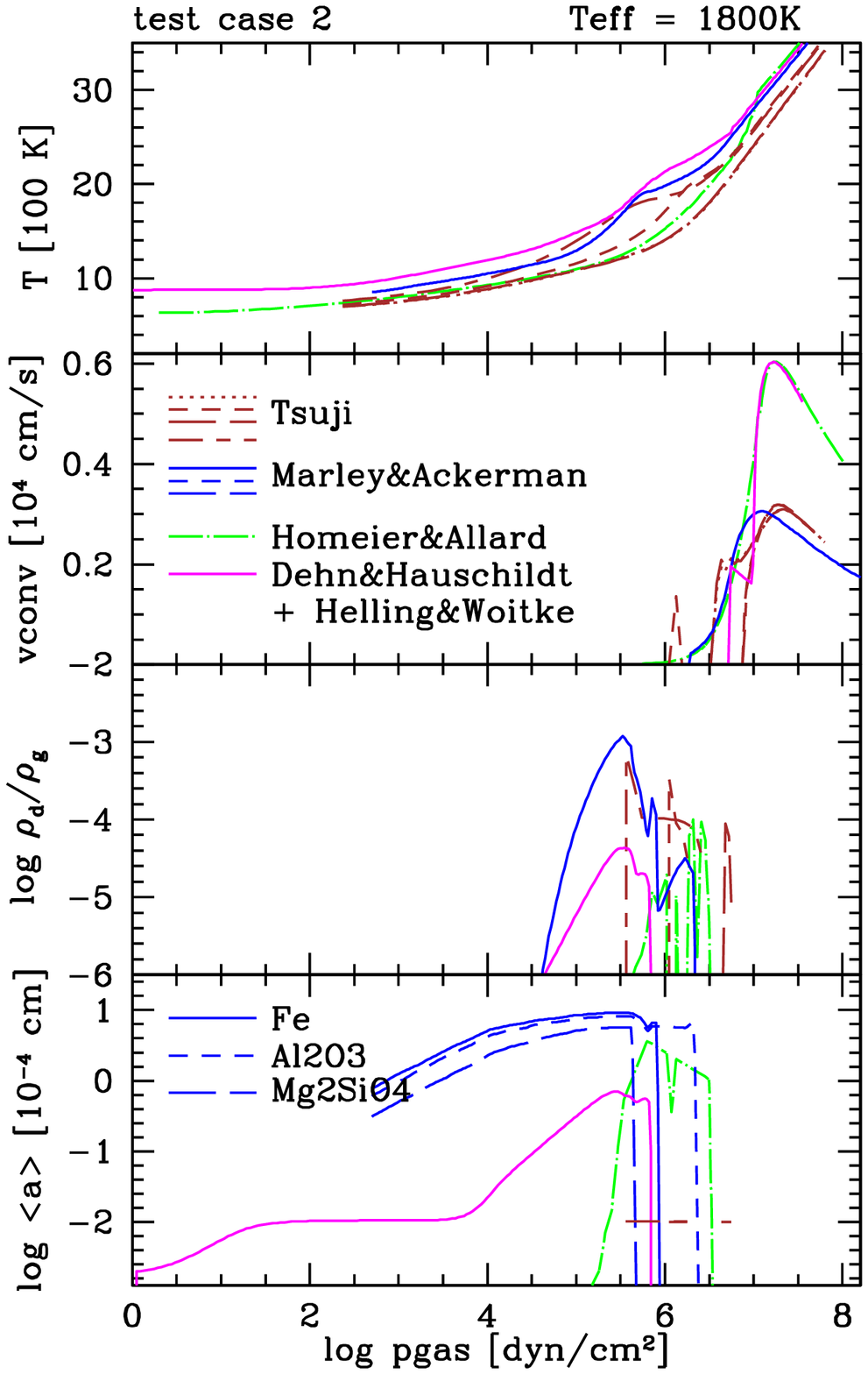}\hspace*{0.2cm}
   \includegraphics[width=9cm]{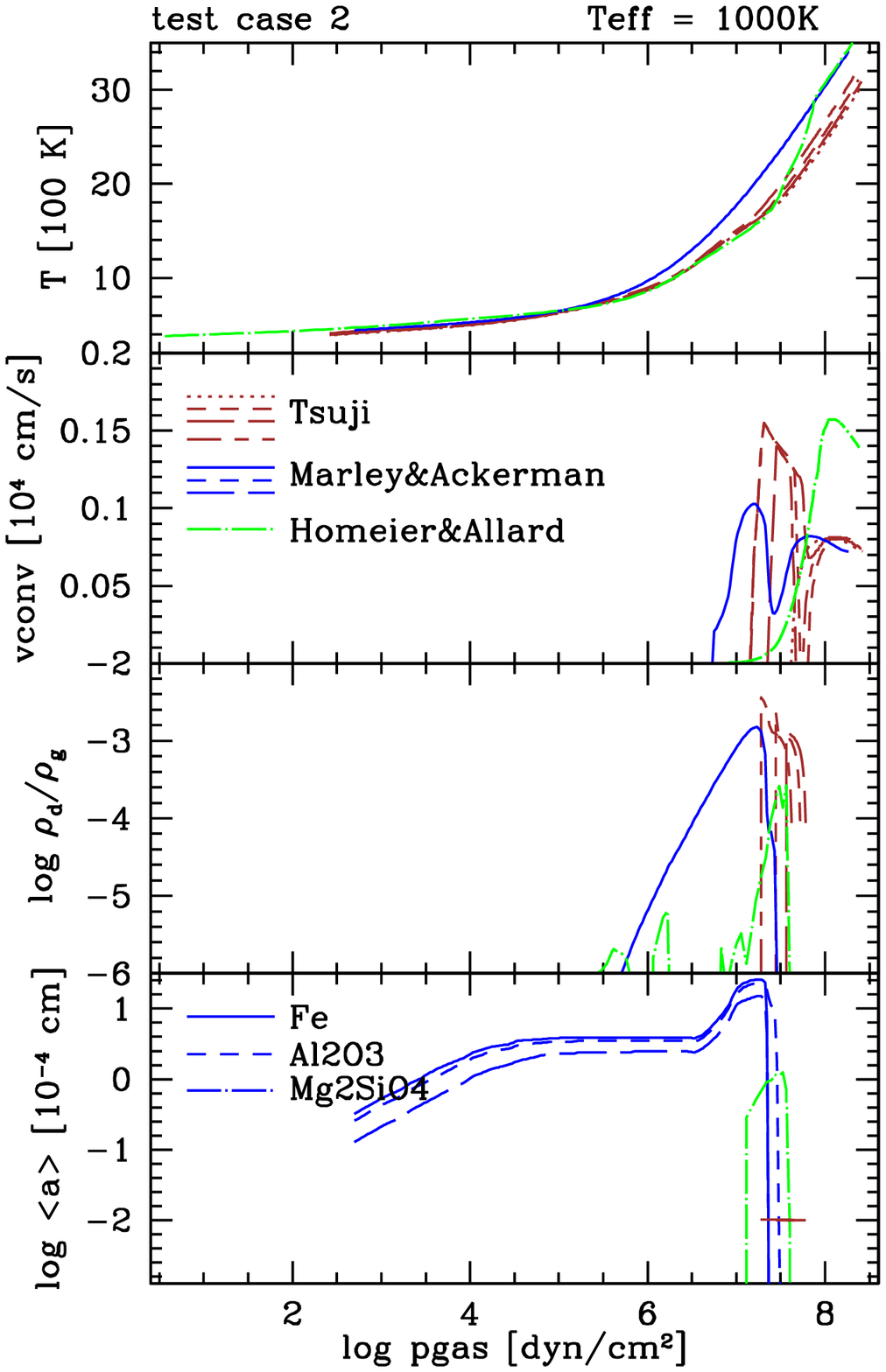}
      \caption{Test cases for complete atmospheric models for
$\log$\,g = 5.0, solar element abundance with T$_{\rm eff}=1800$K
({\bf left}) and T$_{\rm eff}=1000$K ({\bf right}).\newline {\bf
Note:} Different colours stand for different stellar atmospheres
codes. Four models are plotted for the Tsuji-case (brown):
long-short-dashed: T$_{\rm cr}=$1700K (extended cloud), short-dashed:
T$_{\rm cr}=$1800K, long-dashed: T$_{\rm cr}=$1900K (thin cloud),
dotted: no dust opacity considered.  Different line styles in $\log
\langle a \rangle$ indicate different homogeneous dust species in the
Marley, Ackerman \& Lodders-models.}
         \label{fig:P2CloudStruc}
   \end{figure*}

Figure~\ref{fig:P2CloudStruc} shows the $(T, p, v_{\rm
conv})$-profiles and the cloud structures of the complete sub-stellar
atmosphere simulations.

The $(T, p)$-profiles ($1^{\rm st}$ panel) differ considerably in all
parts of the atmosphere. Note that the Tsuji-model without dust
opacity and the Tsuji-model with the highest T$_{\rm cr}=$ 1900K are
almost identical.  The Dehn \& Hauschildt + Helling~\&~Woitke-model is
the hottest throughout the entire atmosphere, where the Tsuji$_{\rm T_{\rm
cr}=1900K}$, and the Tsuji-model without dust opacity, as well as the
Allard \& Homeier-model are the coolest at pressures below $10^{6}$
dyn\,cm$^{-2}$. All models show the backwarming effect at $T\approx
2000$K for T$_{\rm eff}=1800$K except the Allard \& Homeier-model, the
Tsuji$_{\rm T_{\rm cr}=1900K}$ model, and the Tsuji-model without dust
opacity. This backwarming is clearly associated with occurrence of the
cloud layer. It becomes stronger if the maximum amount of dust is
situated at higher altitudes as the comparison of the different
T$_{\rm cr}$-Tsuji-models demonstrates.  The difference to the
Allard \& Homeier-model is understood by noticing that their model
produces much less dust in the respective pressure regime where
$\langle a \rangle \rightarrow 0$ ($3^{\rm rd}$ and $4^{\rm th}$ row
in Fig.~\ref{fig:P2CloudStruc}; also Fig.~\ref{fig:RhoDGAmeanTC1})

The convective velocity $v_{\rm conv}$ explicitly enters all cloud
model but the Tsuji-model ($2^{\rm nd}$ panel). The Allard \& Homeier
and the Dehn \& Hauschildt + Helling \&\,Woitke-model have identical
$v_{\rm conv}$ since they use identical modules for treating the
convective unstable region below the Schwarschild limit but differ in
the treatment of overshooting into the (classically) convective stable
atmosphere at higher altitudes. The Tsuji-model and the
Marley, Ackerman \& Lodders-model are comparable in $v_{\rm conv}$ for the
L-dwarf case T$_{\rm eff}=1800$K.  Both suggest  for the T-dwarf models  a
second convective layer which coincides with maximum $\rho_{\rm
dust}/\rho_{\rm gas}$ in these models (Fig.~\ref{fig:P2CloudStruc},
right).

The dust cloud structures are comparable in the sense that they appear
in approximately the same pressure range except in the Tsuji-model
where the small particles are homogeneously distributed in the cloud
until a certain critical temperature level T$_{\rm crit}$ is
reached. All models produce only the cloud layer of silicates and
oxides\footnote{ The term {\it silicate} is loosely used among
astronomers. In accordance with mineralogy, silicates are those solids
containing Si--O groups. All other solids, like Al$_2$O$_3$[s],
TiO$_2$[s], CaTiO$_3$[s] etc. are oxides.} ($\nearrow$
Sect.~\ref{ss:rhorhog}). The results differ largely in the details
which confirms our results from the first part of our test cases
(Sect.~\ref{s:results1}).  Different vertical cloud extension are
suggested by different simulations: the Allard \& Homeier-model
produces a vertically less extended cloud layer than the Marley,
Ackerman \& Lodders-model and the Dehn \& Hauschildt + Helling\&
Woitke - model, which will have consequence for the emergent spectrum
of such an atmosphere.  It appears that the differences in the cloud
properties are amplified if the entire atmosphere problem is taken
into consideration. The dust-to-gas ratio $\rho_{\rm dust}/\rho_{\rm
gas}$ ($3^{\rm rd}$ row in Fig.~\ref{fig:P2CloudStruc}) differs by 2
orders of magnitudes with the dust-opacity free Tsuji-model providing
the upper limit and the Dehn \&\,Hauschildt + Helling
\&\,Woitke-model, the Allard \& Homeier and the Tsuji$_{\rm T_{\rm
cr}=1700K}$ - models the lower limit. Comparing this with the $(T, p)$-profiles
($1^{\rm st}$ row) suggests that the higher the local temperatures at
a certain pressure level for a given T$_{\rm eff}$=1800K, the smaller
$\rho_{\rm dust}/\rho_{\rm gas}$: $(\rho_{\rm dust}/\rho_{\rm
gas})_{\rm Tsuji_{\rm Tcr}}> (\rho_{\rm dust}/\rho_{\rm gas})_{\rm
Marley, Ackerman \& Lodders}>(\rho_{\rm dust}/\rho_{\rm gas})_{\rm
Dehn \&Hauschildt + Helling \&\,Woitke} $. This trend re-appears for
T$_{\rm eff}$=1000K for which no Dehn \&\,Hauschildt + Helling
\&\,Woitke-model is available.  The Allard \& Homeier-models are
amongst the coolest of the $(T, p)$-profiles for a given T$_{\rm eff}$
but always suggest a smaller $\rho_{\rm dust}/\rho_{\rm gas}$.

  \begin{figure*}
   \includegraphics[width=15.2cm]{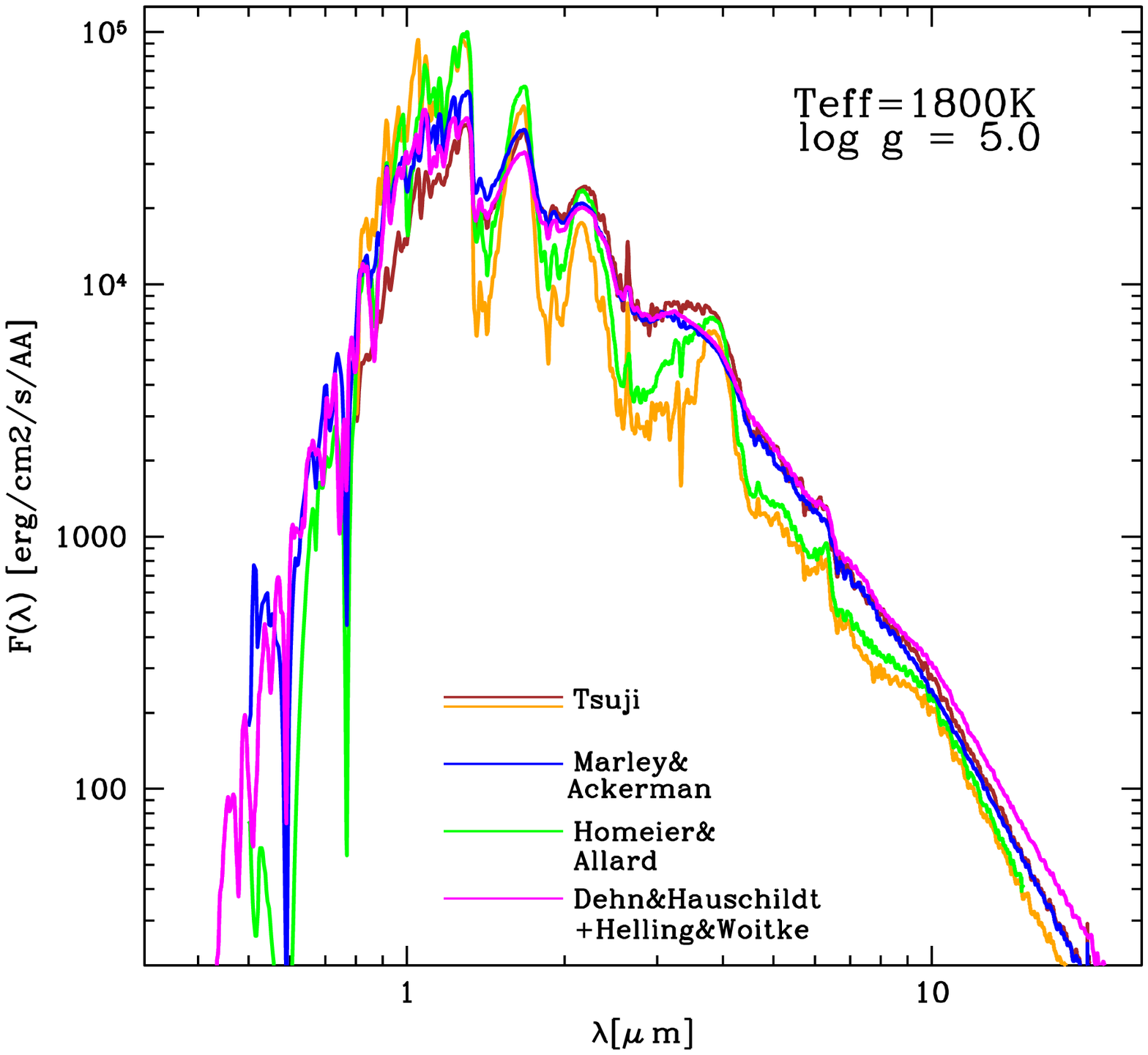}
      \caption{Synthetic spectra for T$_{\rm eff}=$1800K, $\log$g=5.0 and solar metalicity. Two spectra are plotted for the Tsuji-model: T$_{\rm cr}=$1700K (brown; extended cloud) and  T$_{\rm cr}=$1900K (orange; thin cloud).}
         \label{fig:spectotal1800}
   \end{figure*}

  \begin{figure*}
   \includegraphics[width=15.2cm]{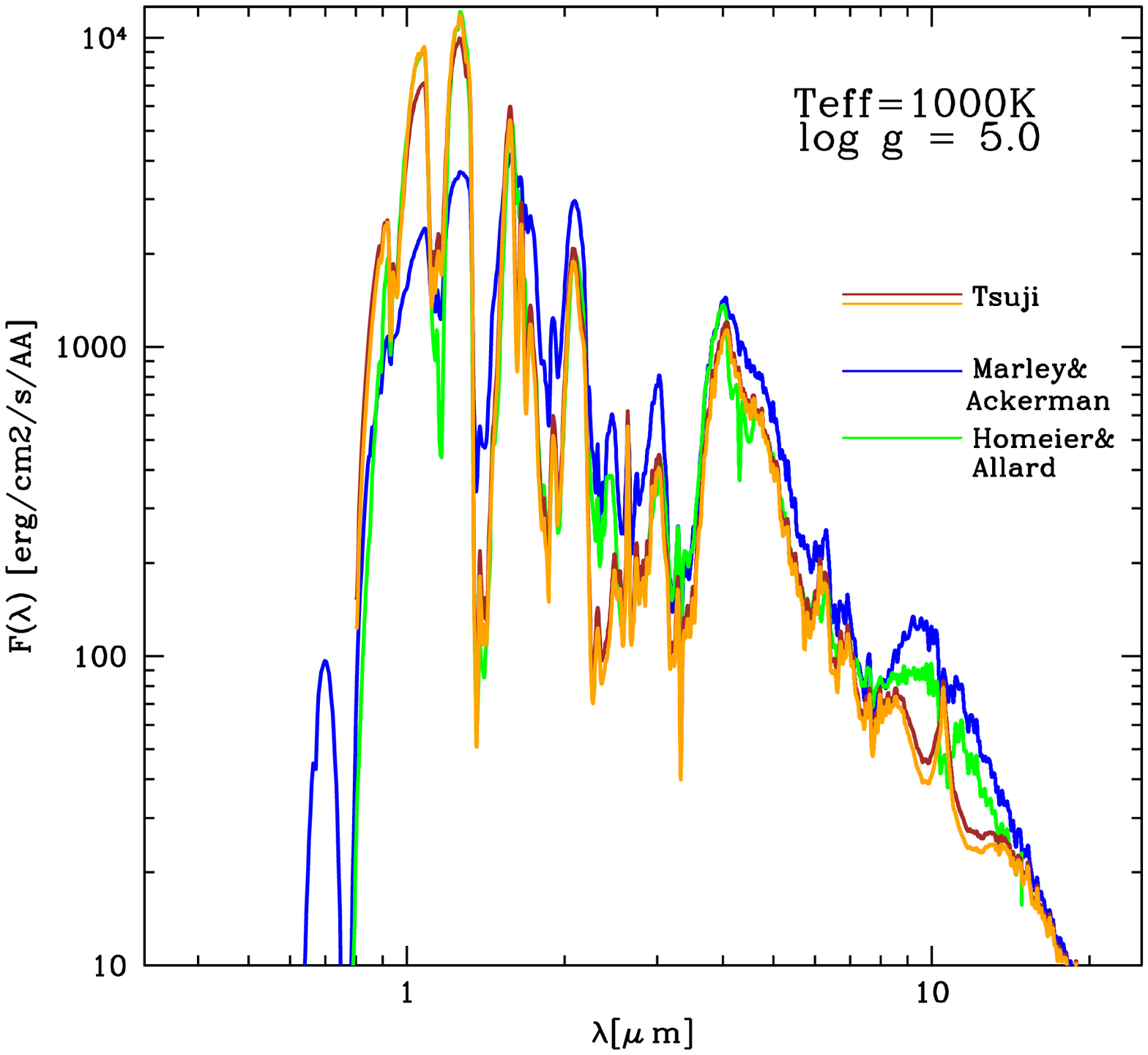}
      \caption{Synthetic spectra for T$_{\rm eff}=$1000K, $\log$g=5.0 and (initial) solar metalicity. Two spectra are plotted for the Tsuji-model: T$_{\rm cr}=$1700K (brown; extended cloud) and  T$_{\rm cr}=$1900K (orange; thin cloud).}
         \label{fig:spectotal1000}
   \end{figure*}

No trend appears in the numerical data for the mean particle sizes
$\langle a \rangle$. The models suggest the biggest particles to
appear at the cloud base except the Tsuji-model which assumes a
constant size in the entire cloud. Note that the particle sizes in the
Allard \& Homeier-models show a very steep distribution towards the
maximum size at the cloud base. It is apparent from
Fig.~\ref{fig:P2CloudStruc} that the different dust cloud treatments
produce different grain size distributions in the atmosphere resulting
in differences in grain sizes up to 2 orders of magnitude at the cloud
top where detectable spectral features would be produced (compare
Marley, Ackerman \& Lodders and Dehn \&\,Hauschildt +
Helling \&\,Woitke). The Allard \& Homeier-models suggest no particles
in these cloud layers, hence, spectral dust features in these models
might be more easily masked by molecular bands.

\begin{table*}
\caption{Photometric fluxes $\log F_{\rm c}$
[erg\,cm$^{-2}$s$^{-1}$\AA$^{-1}$] with c indicating the JHK-2MASS
system, the WFCAM UKIRT system, the VISIR-bands, and the IRAC Spitzer
bands for the models depicted in Fig.~\ref{fig:spec05-3.0_1800}. The
last column contains the maximum differences in $\log F_{\rm c}$
amongst the models: $\Delta_{\rm max}[\log F_{\rm c}]=[\log F_{\rm
c}]^{\rm max}-[\log F_{\rm c}]^{\rm min}$. Note that Tsuji's  thick-cloud case (T$_{\rm cr}=1700$K; each 1st row) is used to calculate $\Delta_{\rm max}[\log F_{\rm c}]$ for the L-dwarf test case (T$_{\rm eff}=1800$K), and Tsuji's  thin-cloud case  (T$_{\rm cr}=1900$K; each 2nd row) for the T-dwarf test case ( T$_{\rm eff}=1000$K). We also list the
photometric fluxes for the HST Vega spectrum of Bohlin \& Gilliland
(2004; Warren 2008, priv. com.) which we use as zero points to calculate the synthetic colours
($\nearrow$ Table~\ref{tab:definitions};
Figs.~\ref{fig:ColCol},~\ref{fig:ColColZYJ}). }
\begin{tabular}{lc|lcc|lcc|cc|cc|cc|c}
\multicolumn{2}{c|}{\bf  Photometric band} & \multicolumn{3}{c|}{} 
                                    & & \multicolumn{2}{c|}{\bf Marley, }
                                    & \multicolumn{2}{c|}{ \bf Allard\&} 
                                    & \multicolumn{2}{c|}{\bf Dehn\&Hauschildt}\\
                &                   & \multicolumn{3}{c|}{\bf Tsuji  }                           
                                    & & \multicolumn{2}{c|}{\bf Ackerman,} 
                                    & \multicolumn{2}{c|}{\bf Homeier}    
                                    & \multicolumn{2}{c|}{\bf +Helling\&Woitke}  &&& {\bf Vega}\\
                &                   & \multicolumn{3}{c|}{ }                           
                                    & & \multicolumn{2}{c|}{\bf \& Lodders} 
                                    & \multicolumn{2}{c|}{ }    
                                    & \multicolumn{2}{c|}{ }\\
\hline 
                &   \hspace*{0.6cm}$\Delta\lambda$  &$T_{\rm cr}$ [K] & \multicolumn{2}{c|}{T$_{\rm eff}$ [K]}
                                    &$f_{\rm sed}$        & \multicolumn{2}{c|}{T$_{\rm eff}$ [K]}
                                    & \multicolumn{2}{c|}{T$_{\rm eff}$ [K]}
                                    & \multicolumn{2}{c|}{T$_{\rm eff}$ [K]}
                                    &\multicolumn{2}{c}{\bf $\Delta_{\rm max}[\log F_{\rm c}]$}
                                    & $\log F_{c0}$\\
   c             &   \hspace*{0.5cm}[$\mu$m]       &     & \hspace*{-0.2cm}1000K &  \hspace*{-0.2cm}1800K
                                    &     & \hspace*{-0.2cm}1000K &  \hspace*{-0.2cm}1800K
                                    & 1000K & \hspace*{-0.2cm}1800K
                                    & 1000K & \hspace*{-0.2cm}1800K
                                    & 1000K &\hspace*{-0.2cm}1800K\\
\hline
 \multicolumn{2}{c|}{\bf JHK-2MASS-bands$^{6}$:} {\bf ($\blacksquare$)} & \multicolumn{10}{c|}{\ }\\
J               & $1.100 - 1.400$       & 1700 & (3.65) & 4.46   & 2 &          & 4.62 &           & 4.76 &      & 4.53 &           & 0.30  & -1.52\\
                 &                                     & 1900 & 3.68   & (4.72) &    & 3.33 &          & 3.65  &          &  --  &          &  0.35 & \\
H              & $1.475 - 1.825$       & 1700 & (3.34) & 4.49   & 2 &          & 4.53  &           & 4.62 &      & 4.45 &           & 0.17 & -1.95\\
                 &                                     & 1900 & 3.28   & (4.51) &   &   3.47 &          & 3.36   &         & --  &           & 0.19  &\\
Ks            & $2.000 - 2.400$       & 1700 & (3.00) & 4.35    & 2 &          & 4.29 &            & 4.29 &      & 4.28 &           & 0.07  & -2.37\\
                &                                     & 1900 & 2.95    & (4.14) &     & 3.23 &          & 3.02   &          &  --  &          & 0.28 &\\
\hline
 \multicolumn{2}{c|}{\bf ZYJHK-UKIRT-bands$^{7}$:} & \multicolumn{10}{c|}{\ }\\
Z               & $0.830 - 0.925$       & 1700 & (3.26) & 3.87   & 2 &          & 4.22 &           & 4.17 &     & 4.14 &          & 0.35 & -1.07\\
                &                                      & 1900 & 3.22   & (4.39) &    & 2.85  &          & 2.94  &         &  --  &         & 0.37 & &     \\
Y               & $0.970 - 1.070$       & 1700 & (3.74) & 4.31   & 2 &          & 4.50 &          & 4.59 &       & 4.56  &         & 0.28 & -1.24\\
                &                                      & 1900 & 3.84    & (4.78) &    & 3.29 &          & 3.84 &          & --   &           & 0.55& &     \\
J               & $1.170 - 1.330$       & 1700 & (3.86) & 4.55  & 2 &            & 4.69 &          & 4.89 &      & 4.61 &          & 0.34 & -1.53\\
                &                                      & 1900 & 3.90   & (4.86)&    & 3.49   &          & 3.89 &          & --   &         &  0.41& &     \\
H               & $1.490 - 1.780$      & 1700 & (3.34) & 4.49   & 2 &           & 4.52 &           & 4.61 &      & 4.44 &         & 0.17 & -1.94\\
                &                                      & 1900 & 3.28   & (4.50) &    & 3.45   &         & 3.36  &          & --  &          & 0.17 & &     \\
K               & $2.030 - 2.370$       & 1700 & (2.90) & 4.34   & 2 &            & 4.27&           & 4.27 &      & 4.26 &         &0.08 & -2.41\\
                &                                      & 1900 & 2.85    & (4.11) &    & 3.14   &         & 2.94  &          & --   &         & 0.29& &     \\
\hline                                
 \multicolumn{2}{c|}{\bf IRAC Spitzer-bands$^{8}$: ($\blacktriangle$)}  & \multicolumn{10}{c|}{\ }\\
Band~1          & $2.965 - 4.165$   & 1700 & (2.56) & 3.90   & 2 &          & 3.81 &           & 3.80 &       & 3.83 &          &  0.10 & -3.19\\
                &     [3.6]                            & 1900 & 2.52    & (3.66) &    & 2.69 &          & 2.70  &          & --   &          &  0.18 \\
Band~2          & $3.704 - 5.324$   & 1700 & (2.85) & 3.53    & 2 &          &3.48  &           & 3.36 &       & 3.52 &           & 0.17 & -3.58\\
                &     [4.5]                            & 1900 & 2.83    & (3.28) &     & 2.97 &          & 2.80  &          & --    &          & 0.17 & \\
Band~3          & $4.626 - 6.896$   & 1700 & (2.29) & 3.18    & 2 &          &3.15  &            & 2.99 &       & 3.20 &          & 0.21 & -3.99\\
                &     [5.8]                            & 1900 & 2.26    & (2.93) &     & 2.45 &          & 2.29   &          & --    &          & 0.19 & \\
Band~4          & $5.618 - 10.31$   & 1700 & (1.88) & 2.72    & 2 &           &2.71  &            & 2.57 &       & 2.76 &          & 0.19 & -4.51\\
                &     [8.0]                            & 1900 & 1.85    & (2.50) &     & 2.02  &          & 1.95   &          &  --  &           & 0.17\\
\hline
 \multicolumn{2}{c|}{\bf VISIR-bands$^{9}$:}  & \multicolumn{10}{c|}{\ }\\
PAH1            & $8.38 - 8.8$      & 1700 & (1.87) & 2.61   & 2 &          & 2.59 &          & 2.49 &       & 2.64 &          & 0.15  & -4.69\\
                &                                   & 1900 & 1.84    & (2.43) &    & 2.09 &          & 1.93 &          &  --   &          & 0.25 &\\
ArIII           & $8.92 - 9.06$     & 1700 & (1.80) & 2.57  & 2 &          & 2.54 &           & 2.46 &        & 2.60    &           & 0.14 & -4.76 \\
                &                                & 1900 & 1.76   & (2.41) &   & 2.06  &          & 1.94  &         & --     &             &  0.30  & \\
SIV             & $10.410 - 10.570$ & 1700 & (1.85) & 2.36   & 2 &           & 2.32 &          & 2.26 &       & 2.43 &         & 0.17  & -5.03\\
                &                                       & 1900 & 1.82   & (2.23) &    & 1.89  &          & 1.65 &          & --   &          & 0.24 &\\
PAH2            & $10.965 - 11.555$ & 1700 & (1.52) & 2.24    & 2 &           & 2.20 &           & 2.16 &       & 2.31 &         & 0.15  & -5.15\\
                &                                          & 1900 & 1.46    & (2.11) &    &  1.89 &           & 1.74 &           & --   &          & 0.43 & \\ 
SiC             & $10.680 - 13.020$ & 1700 & (1.51) & 2.17    & 2 &         & 2.13 &            & 2.09 &      & 2.25 &          & 0.16  & -5.22\\
                &                                       & 1900 & 1.46    & (2.04) &    & 1.81&           & 1.66   &          & --  &          & 0.35 &\\
NeII            & $12.695 - 12.905$ & 1700 & (1.43) & 2.02   & 2 &          & 1.98 &           & 1.92 &     & 2.10 &           & 0.18  & -5.37\\
                &                                       & 1900 & 1.38   & (1.88) &    & 1.66 &           & 1.53 &          & --  &          & 0.28 & \\
Q1              & $17.235 - 18.065$ & 1700 & (1.03) & 1.44   & 2 &          & 1.43 &      & --   &      & 1.59 &           &  0.15 & -5.93\\
                &                                       & 1900 & 1.02   & (1.31) &    & 1.06 &          &  --  &      &  --  &          & 0.04 & \\
Q2              & $18.280 - 19.160$ & 1700 & (0.94) & 1.34   & 2 &          & 1.33 &       & --   &       & 1.50 &           & 0.17 & -6.03\\
                &                                       & 1900 & 0.93   & (1.22) &    & 0.97 &          &  --  &       & --   &           & 0.04 &\\
Q3              & $19.300 - 19.700$ & 1700 & (0.85) & 1.28   & 2 &          &1.27  &      & --   &       & 1.44 &           & 0.17  & -6.10\\
                &                                       & 1900 & 0.84   & (1.15) &    & 0.88 &          &  --  &      & --   &           & 0.04 &  \\
\end{tabular}
\label{tab:photFlux}
\end{table*}%

\subsubsection{Spectral energy distribution}\label{ss:specs}

\begin{figure*}
   \includegraphics[width=8.5cm]{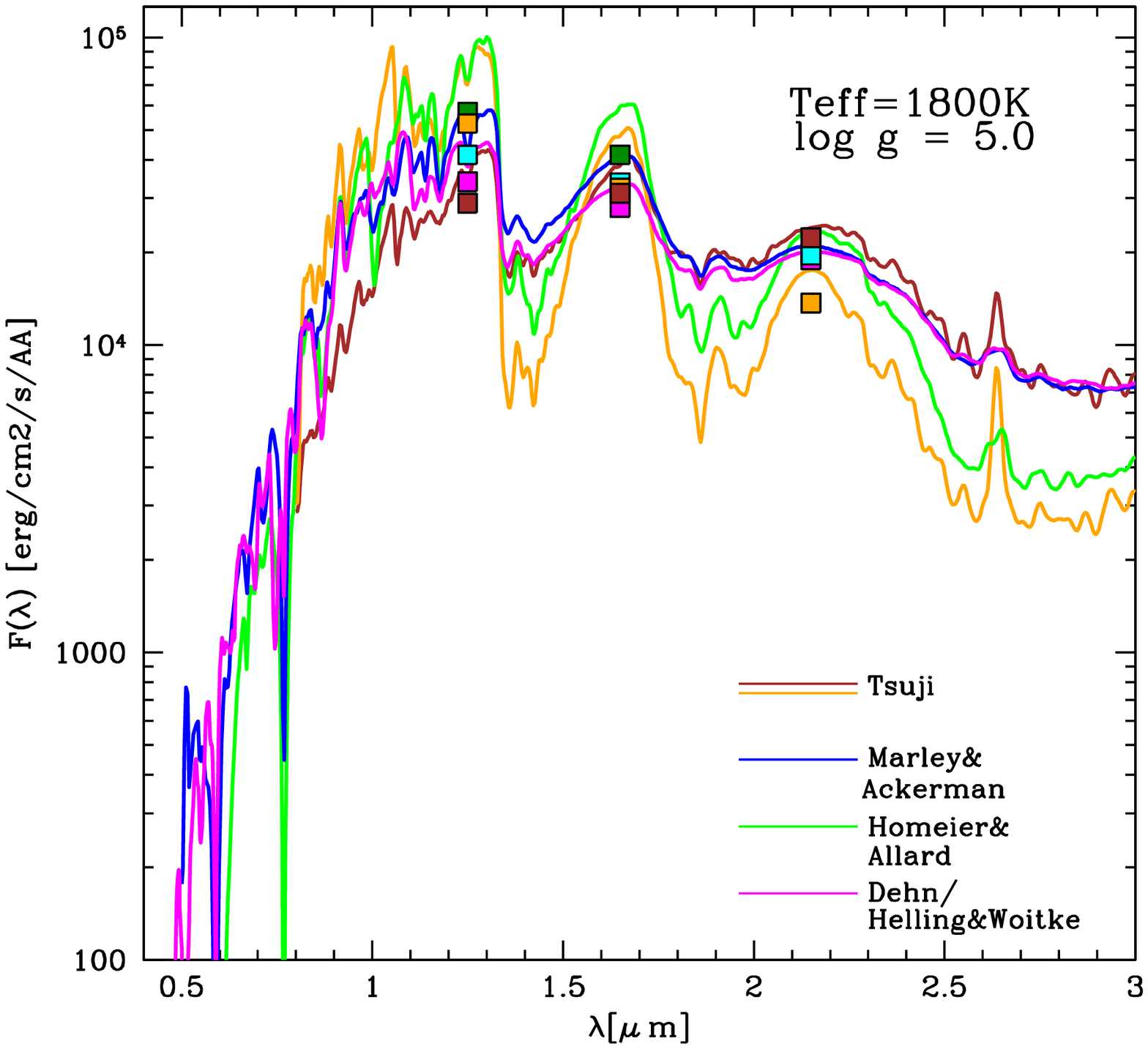}
   \includegraphics[width=8.5cm]{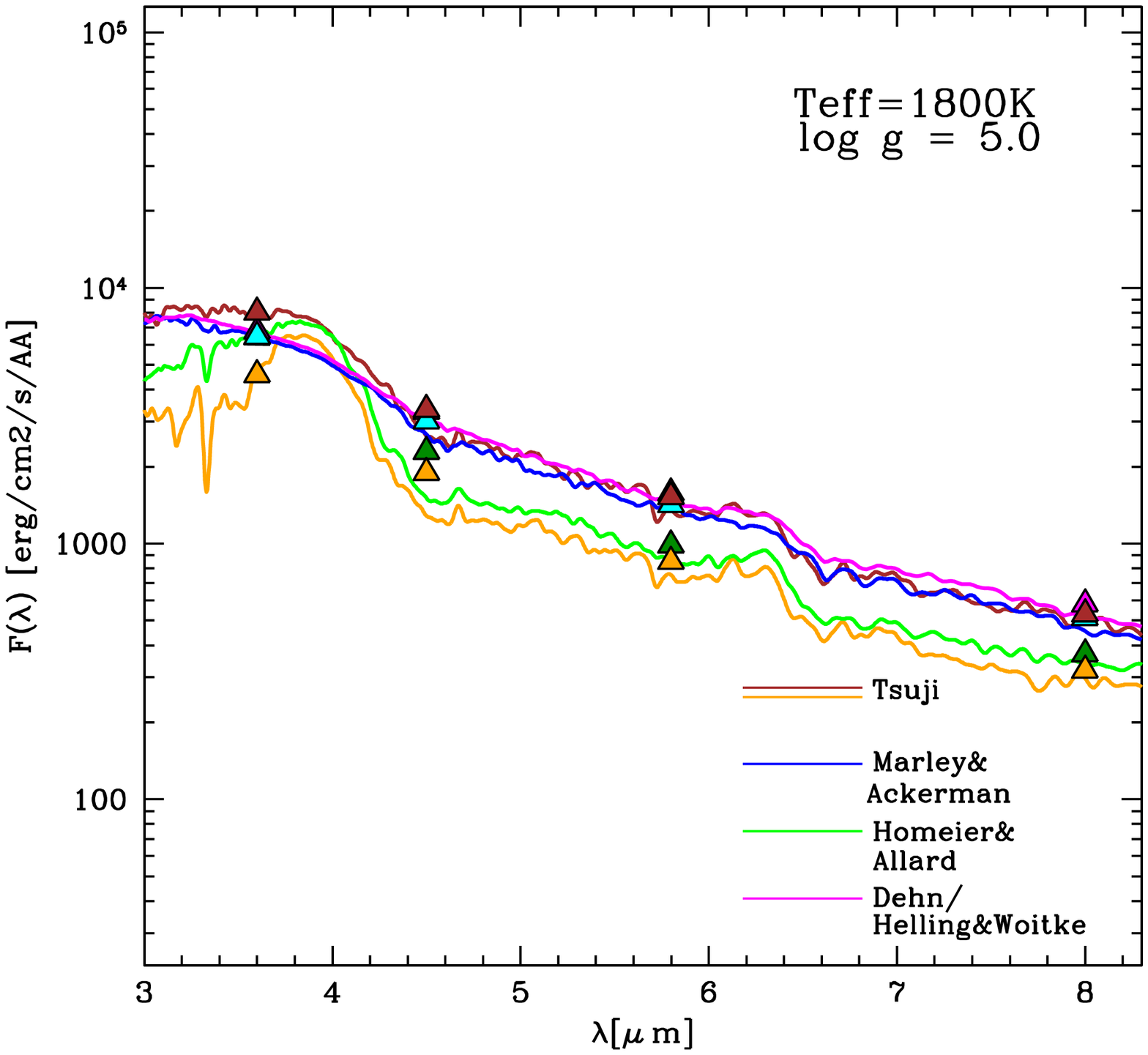}\\
   \includegraphics[width=8.5cm]{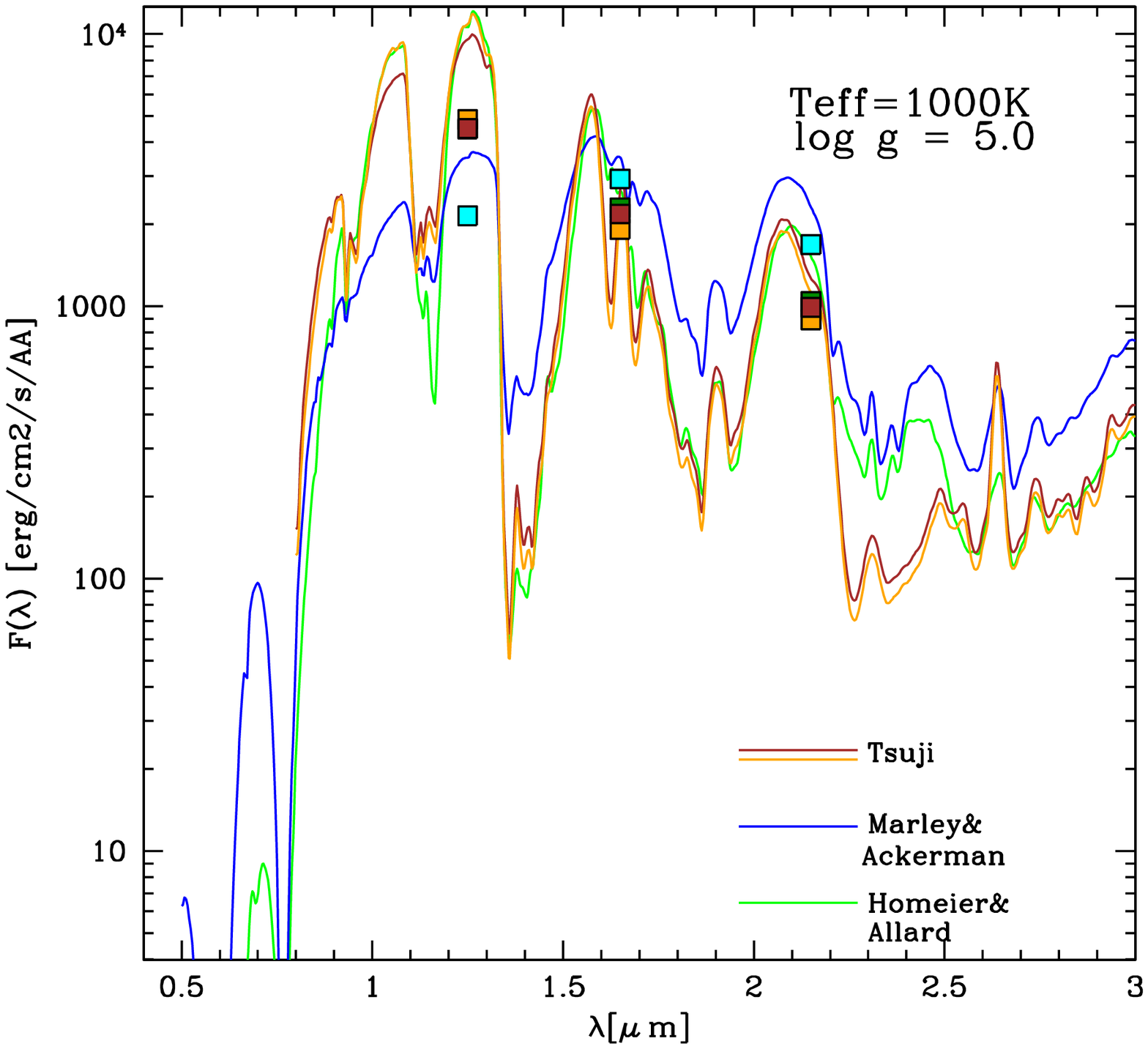}
   \includegraphics[width=8.5cm]{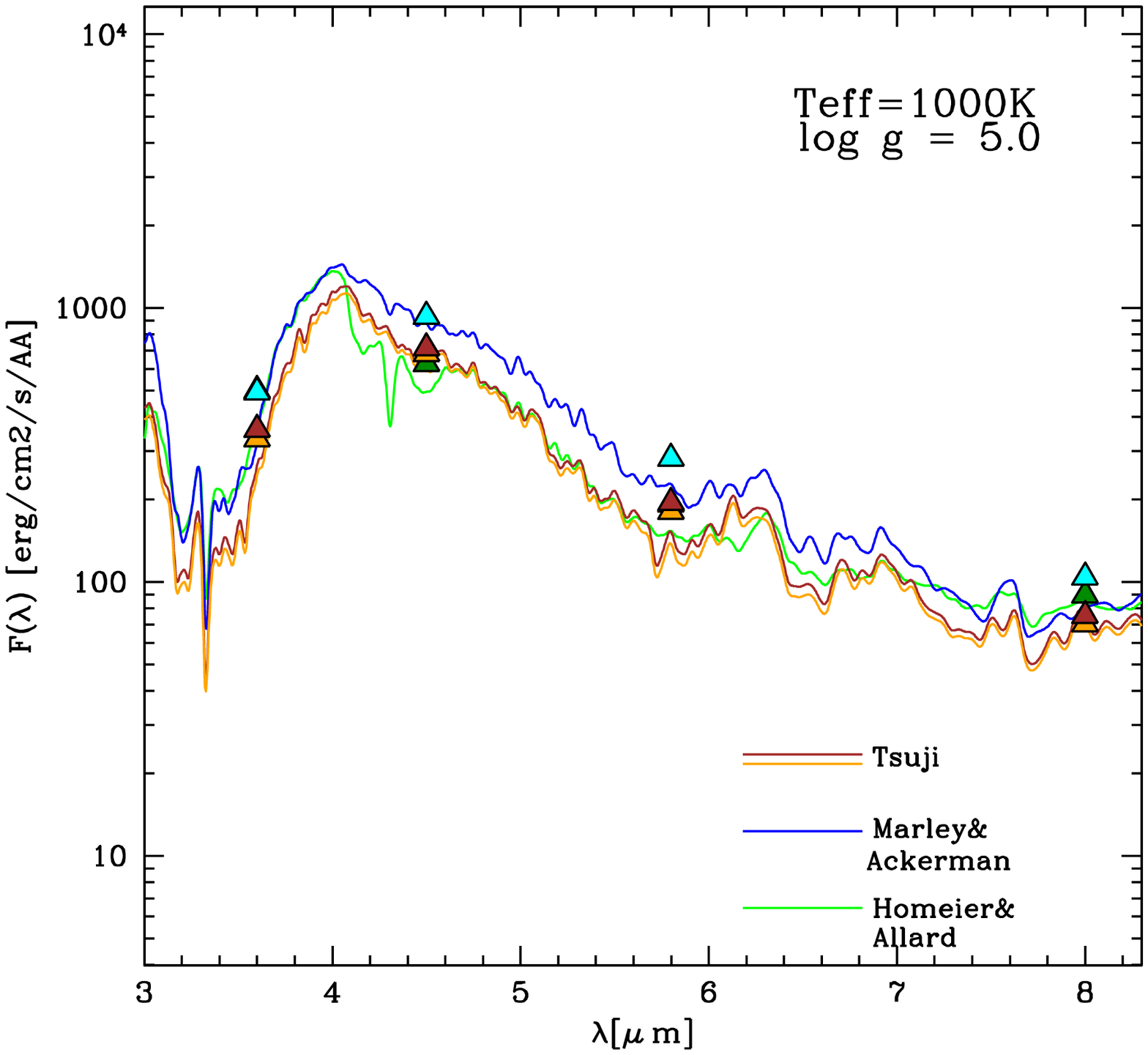}
      \caption{Synthetic spectra for T$_{\rm eff}=$1800K ({\bf top})
       and T$_{\rm eff}=$1000K ({\bf bottom}) with $\log$g=5.0 and solar
       metalicity for a spectral resolution R=200. Two spectra are
       plotted for the Tsuji dust model: T$_{\rm cr}=$1700K (brown; extended cloud)
       and T$_{\rm cr}=$1900K (orange; thin cloud). Photometric fluxes (symbols)
       are plotted for the JHK-2MASS-system (left panels,
       $\blacksquare$) and the IRAC Spitzer-bands (right panels,
       $\blacktriangle$) at the band center frequency
       $\Delta\lambda/2$. The photometric fluxes are summarised in
       Table~\ref{tab:photFlux}.}
         \label{fig:spec05-3.0_1800}
   \end{figure*}

The spectral energy distributions (SEDs) between $0.5\,\ldots\,18\mu$m
calculated by the different model atmosphere codes employing different
cloud models are depicted in Fig.~\ref{fig:spectotal1800} (T$_{\rm
eff}=$1800K) and Fig.~\ref{fig:spectotal1000} (T$_{\rm
eff}=$1000K). The Marley, Ackerman \& Lodders model for $T_{\rm eff} =
1000\,\rm K$ employed $f_{\rm sed}=2$ for consistency with the $T_{\rm
eff} = 1800 \,\rm K$ case.  Modeling by this group suggests
that the spectra of early T dwarfs are better fit with larger values
of $f_ {\rm sed}$.  The thick clouds resulting from the choice of
$f_{\rm sed}=2$ (Fig. 5, bottom right) are responsible for the shallow
absorption bands and red colors compared to the other groups for this
case. In the Tsuji-case we plot two models, T$_{\rm cr}=1700$K and
T$_{\rm cr}=1900$K which demonstrate a very thin (T$_{\rm cr}=1900$K)
and an extended (T$_{\rm cr}=1700$K) cloud layer.  No two codes
produce identical SEDs. Generally, Allard \& Homeier and the
Tsuji$_{\rm Tcr=1900K}$-models appear brighter than the other models
between $\sim 0.8\,\ldots\sim 1.5\mu$m in the optical and near-IR (see
also left panels Fig.~\ref{fig:spec05-3.0_1800}). The
Dehn~\&\,Hauschildt + Helling~\&\,Woitke, the Marley, Ackerman \&
Lodders and the Tsuji$_{\rm Tcr=1700K}$-models are the brightest of
all models T$_{\rm eff}=$1800K in the IR for $\lambda> 5\mu$m. This
result is not surprising because the Dehn~\&\,Hauschildt +
Helling~\&\,Woitke, the Marley, Ackerman \& Lodders and the
Tsuji$_{\rm Tcr=1700K}$-models contain the highest amount of small
dust particles in the upper cloud layers (Sect.~\ref{ss:atmprof}) and
should therefore produce a redder atmosphere compared to a model
without dust at comparable atmosphere pressures. In principle, the
same analysis applies for the T$_{\rm eff}=$1000K-case, representing
the T-dwarf regime within this study
(Fig.~\ref{fig:spectotal1000}). The cloud has moved already
considerably below $\tau=1$ that both Tsuji-models appear very
similar. Additionally, the Allard \& Homeier-model suggests more
spectral flux in several wavelength intervals than the Tsuji$_{\rm
Tcr=1900K}$-model with an extended cloud
layer. Figure~\ref{fig:spectotal1000} also demonstrates an appreciable
difference of the Tsuji-models around $10\mu$m which are sensitive to
the application of the JOLA opacity band model.

It is apparent from Fig.~\ref{fig:spec05-3.0_1800} that the
Tsuji$_{\rm Tcr=1900K}$, the Marley, Ackerman \& Lodders-models and the
Dehn~\&\,Hauschildt + Helling~\&\,Woitke-model produce much shallower
absorption features in particular in the optical and near-IR then the
Tsuji$_{\rm Tcr=1700K}$ and the Allard \& Homeier-models. Again, the reason lies in the
differences in cloud modelling. This result appears surprising for the
Dehn~\&\,Hauschildt + Helling~\&\,Woitke~model since more elements
remain in the gas phase due to incomplete dust formation, hence, the
absorption features should be deeper. A comparison with
Fig.~\ref{fig:P2CloudStruc} (top panel) resolves this: All models
showing shallow absorption features are amongst the hottest $(T,p)$
structures, hence, have low densities at given atmospheric temperature
and therefore lower opacity at that atmospheric height. Note that the
$(T, p)$-profiles are much more similar for T$_{\rm eff}=$1000K and so
is the depth of the absorption features.  Consequently, a spectral
analysis relying on the depth of near- and IR spectral features,
e.g. as gravity indicator, would underestimate the gravity in the case
of the Tsuji and the Allard \& Homeier-models compared to the
Marley, Ackerman \& Lodders and Dehn~\&\,Hauschildt + Helling~\&\,Woitke~models
in the L\,-\,dwarf regime.

It appears that the treatment of the gas-phase opacity can
account for some differences in the synthetic spectral energy
distribution, here in particular the treatment by the JOLA band method
in the Tsuji-models vs. the more frequency sensitive methods used in all other
models. However, the completeness of the molecular line lists has only
minor effects on our photometry results given the large influence of
the dust modelling demonstrated here.

\subsubsection{Photometric fluxes}

\begin{figure}
   \includegraphics[width=8.8cm]{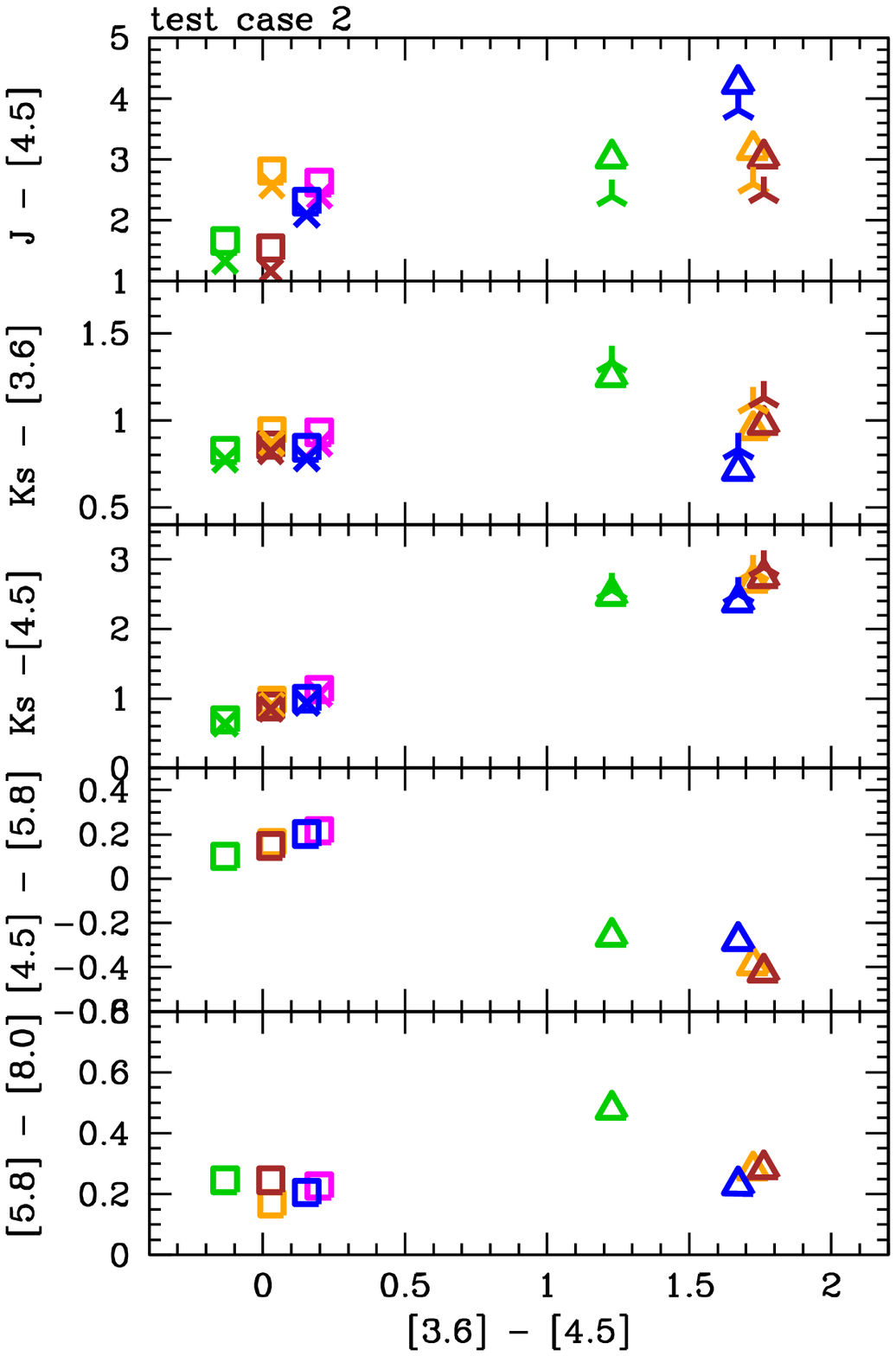}
      \caption{Synthetic colour-colour diagrams for 2MASS, for T$_{\rm eff}=$1800K
      ($\Box$ and cross) and T$_{\rm eff}=$1000K ($\triangle$ and star) with
      $\log$g=5.0 and solar metalicity. The symbol-colour code is the
      same like in Fig.~\ref{fig:spectotal1800}:
      Tsuji T$_{\rm cr}=$1700K (brown, thick cloud) and T$_{\rm cr}=$1900K
      (orange, thin cloud), Marley, Ackerman \& Lodders (blue), Allard \& Homeier
      (green), Dehn\& Hauschildt + Helling \& Woitke
      (magenta). The star symbols  indicate the colours evaluated for the WFCAM UKIRT filter system. }
         \label{fig:ColCol}
   \end{figure}

\begin{figure}
  \centering
 \includegraphics[width=8.1cm]{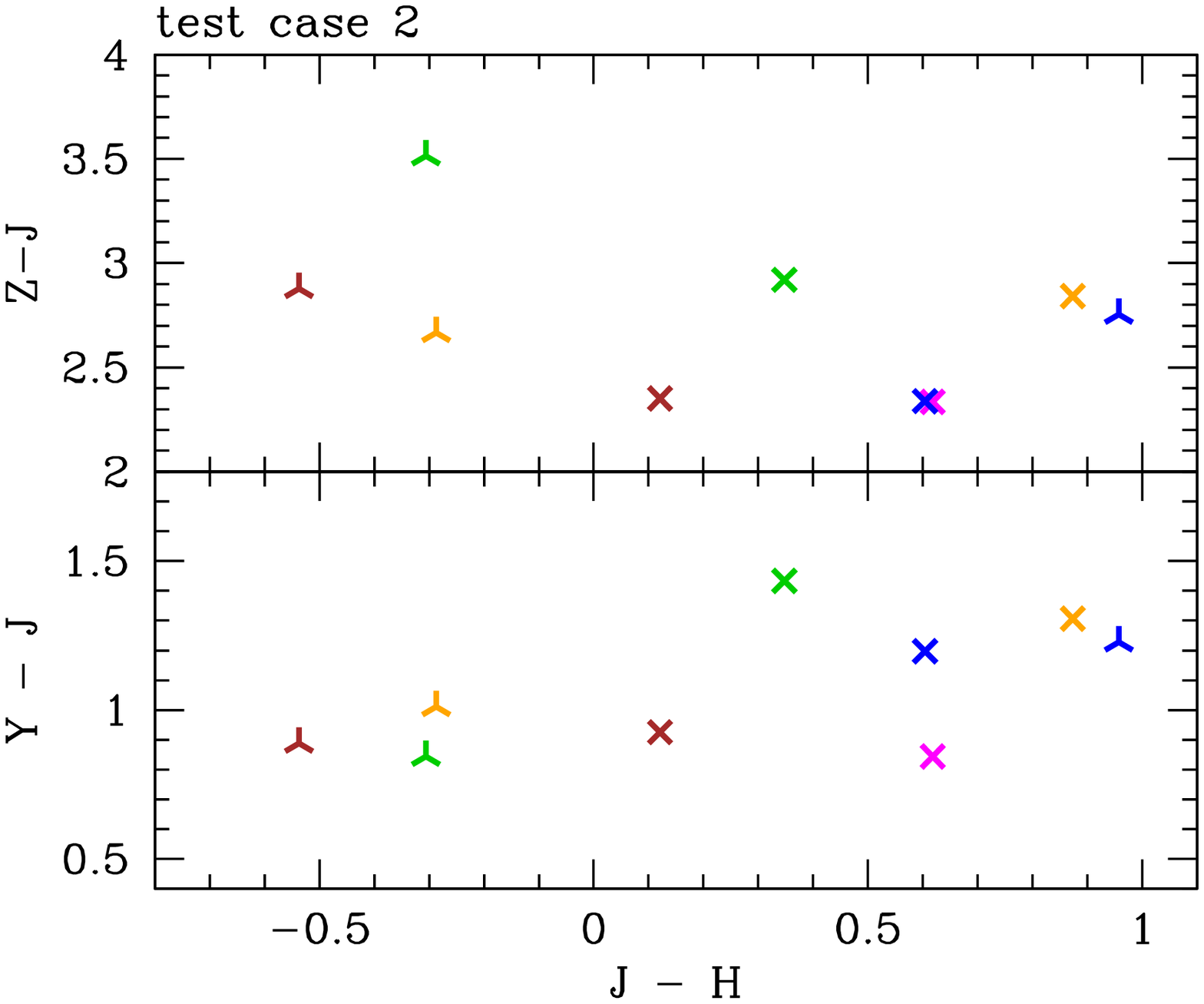}
      \caption{Synthetic colour-colour diagrams only for WFCAM UKIRT filters 
      for T$_{\rm eff}=$1800K (cross) and T$_{\rm eff}=$1000K (star)
      with $\log$g=5.0 and solar metalicity. The symbol-colour code is
      the same like in Fig.~\ref{fig:spectotal1800}: Tsuji T$_{\rm
      cr}=$1700K (brown, thick cloud) and T$_{\rm cr}=$1900K (orange, thin cloud), Marley, Ackerman \& Lodders (blue),  Allard \& Homeier (green), Dehn\&
      Hauschildt + Helling \& Woitke (magenta).}
         \label{fig:ColColZYJ}
   \end{figure}

We wish to compare our simulations also in terms of photometric fluxes
(Figs.~\ref{fig:spec05-3.0_1800},~\ref{fig:ColCol},~\ref{fig:ColColZYJ}
and Table~\ref{tab:photFlux}). We have chosen to demonstrate our
comparison for four filter systems covering the near-IR and the IR:
the JHK-2MASS photometric
system\footnote{http://web.ipac.caltech.edu/staff/waw/2mass/opt$_{-}$cal/index.html},
the WFCAM UKIRT
filters\footnote{http://www.ukidss.org/technical/technical.html}, the
IRAC Spitzer photometric
bands\footnote{http://ssc.spitzer.caltech.edu/irac/spectral$_{-}$response.html},
and the VISIR VLT
system\footnote{http://www.eso.org/sci/facilities/paranal/instruments/visir/inst/index.html}.
Delfosse et al. (2000) show for their empirical mass-luminosity
relation that model atmosphere results are more reliable in the
near-IR (JHK) than at lower wavelength. However, we include the Z and
Y (+JHK) from WFCAM UKIRT filter system for comparison. Note that
Carpenter (2001) provides transformation formula for the 2MASS colours
into a number of different photometric systems (also Hewett et
al. 2006).

The Allard \& Homeier-model exhibits the highest J, H and Y
photometric fluxes in the $T_{\rm eff}=1800$K-case,
while both Tsuji-models bracket the compared models for the IRAC
photometric fluxes (except Band 4). The Allard \& Homeier, and the
Tsuji$_{\rm T_{\rm cr}=1900K}$-models have the largest Y fluxes in the
$T_{\rm eff}=1000$K-case, and the Tsuji$_{\rm T_{\rm cr}=1700K}$-model
has the largest flux in Z. The Marley, Ackerman \& Lodders-model and the Tsuji$_{\rm T_{\rm cr}=1900K}$-model
suggest the largest flux in Z for $T_{\rm eff}=1800$K. Note that
$\log F_{\rm J}^{\rm UKIRT}> \log F_{\rm J}^{\rm 2MASS}$, $\log F_{\rm
H}^{\rm UKIRT}\approx \log F_{\rm H}^{\rm 2MASS}$, and $\log F_{\rm
K}^{\rm UKIRT}< \log F_{\rm K}^{\rm 2MASS}$ for all model
approaches. The model results differ the most in the UKIRT ZYJ bands in both test cases.

Interestingly, the Tsuji models suggest the faintest fluxes in all
IRAC band for $T_{\rm eff}=1000$K while the Marley, Ackerman \&
Lodders-models result in the largest fluxes in these wavelength
bands. The maximum difference amongst the models in photometric fluxes
($\Delta_{\rm max}[\log F_c]$, last columns Table~\ref{tab:photFlux})
are larger in the T-dwarf test case than in the L-dwarf test case.
The maximum differences occur in the ZYJH bands, and in the VISIR SIV,
SiC, and NeII bands for the T-dwarf test case.

These synthetic photometry allows to suggest an error margin due to
spread in the model results for apparent magnitudes (definition see
Table~\ref{tab:definitions}).  {\it The photometric flux differences
relate to uncertainties in apparent magnitudes between $0.25 < \Delta
m < 0.875$ for the L-dwarf test case. The uncertainty in apparent
magnitudes increases considerably for the T-dwarf test case due to
strong differences in the Y band: $0.1 < \Delta m < 1.375$} .

Figures~\ref{fig:ColCol} and ~\ref{fig:ColColZYJ} demonstrate how the
photometric fluxes of the model atmosphere codes translate into
colours (definition see Table~\ref{tab:definitions}). For this, we
have normalised the photometric fluxes to the corresponding
photometric fluxes for Vega ($\log F_{\rm c0}$ in
Tables~\ref{tab:definitions} and~\ref{tab:photFlux}). The largest differences occur for the
L-dwarf model (T$_{\rm eff}=1800$K) in J-[4.5], Ks-[4.5], K-[4.5], and the
[3.5]-[4.5].  For the T-dwarf model (T$_{\rm eff}=1000$K) the maximal differences occur in 
Ks-[3.6], [3.5]-[4.5], and Ks-[4.5].

 If we assume the synthetic colours are correct, we can in principle
use them to infer the spectral type of an object from an observed
(colour, SpT)-diagram\footnote{SpT = Spectral Type}. Ideally, the
synthetic colours derived in Fig.~\ref{fig:ColCol} for T$_{\rm
eff}$=1800K should result in a L-dwarf and the colours for T$_{\rm
eff}$=1000K should suggest a T-dwarf.  For this exercise, the
synthetic colours ($m_1 - m_2$) in Fig.~\ref{fig:ColCol} are compared
with the observed IRAC colours in Patten et al. (2006), and the
related spectral type is picked from their Fig.~10.  However, in
Patten et al. (2006) the [3.6]-[4.5]--SpT relation appears ambiguous
for M7$\,\ldots\,$L8 since [3.6]-[4.5] remain approximately constant
for these spectral types. Our synthetic error margin is slightly
larger than the [3.6]-[4.5] scatter in Patten et al. (2006). We
encounter a similar challenge in relating our synthetic mean value for
[4.5]-[5.8] to a possible spectral type suggesting M0$\,\ldots\,$L7 in
the worst case, but the mean value for [4.5]-[5.8] results in SpT=
M8$\,\ldots\,$L6. Patten et al. (2006) argue that the scatter in their
observed [5.8]-[8.0]--SpT plot might be due to the H$_2$O vs. CH$_4$
absorption at 7.7$\mu$m. Our synthetic mean value would then suggest
with these observed data an interval of M9$\,\ldots\,$T5 inside the
synthetic error bars, while the mean value for [5.8]-[8.0] narrows the
SpT range to L0$\,\ldots\,$T4. About the same conservative spectral
type range is found for J-[4.5], but can be narrowed to
L3$\,\ldots\,$L5 for the J-[4.5]-mean value. For our T-dwarf test case
(T$_{\rm eff}=1000$K) the biggest uncertainties in comparison with the
Patten et al (2006) data occur for [5.8]-[8.0], Ks-[3.6] and J-[4.5].
However, the SpT-colour relation in Patten et al. (2006) is much
narrower for T-dwarfs, and hence, our synthetic mean colours do
suggest much narrower SpT ranges than for our L-dwarf test case
(compare Table~\ref{tab:synSpecType}).

Interestingly, the two extreme Tsuji-models (T$_{\rm cr}=1700$K and
T$_{\rm cr}=1900$K) bracket the Y-J and the Z-J UKIRT colours in
Fig~\ref{fig:ColColZYJ}.  We compare our synthetic Y-J and J-H UKRIT
colours to Hewett et al. (2006), and we reproduce the spectral class
of our test case models better than for the near-IR colours (see
Table~\ref{tab:synSpecType}). A comparison with Lodieu et al.(2007a)
demonstrates that synthetic Z-J colours falls well in their sequence
of substellar objects with decreasing mass. However, a proper
reproduction of a T-dwarf spectrum would demand an adjustment of cloud
parameters like T$_{\rm crit}$, $f_{\rm sed}$ and possibly also the mixing
efficiency as described in Sections~\ref{tsuji:cloudmodel},
~\ref{HomAll:cloudmodel},~\ref{Marely:cloudmodel}
and~\ref{HeWo:cloudmodel} compared to the parameter used in this
comparison study which are more suitable for L-dwarf model atmospheres
(see e.g. Tsuji 2005, Cushing et al. 2008, Stephens et al. 2008).

\begin{table}
\centering
\caption{Mean synthetic colours (${(m_1 - m_2)}_{\rm mean}= \sum^{L}_l
(m_1 - m_2)_l /L$; L -- total number of models) and error margins 
($((m_1 - m_2)_{\rm max} - (m_1 - m_2)_{\rm min})/2$ - maximum colour difference) derived from
Figs.\ref{fig:ColCol},~\ref{fig:ColColZYJ}. The HST Vega spectrum of
Bohlin \& Gilliland (2004; Warren 2008, priv. com.) is used as zero
point. Listed are also the spectral types suggested by the mean synthetic colours alone (each 1st row) and including the synthetic error margin (each 2nd row).}
\begin{tabular}{clcc}
\multicolumn{3}{l}{{\bf L-dwarf test case (5 models):}}\\
colour          & \multicolumn{3}{c}{${(m_1 - m_2)}_{\rm mean}^{1800K}\,\, \Rightarrow$ \,\,SpT$^{10}$} \\
\hline
$ [3.6]-[4.5]$   & $0.0558                 $  &$\Rightarrow$&  M7$\,\ldots\,$L7\\ 
                         & $0.0558\pm 0.175$  & $\Rightarrow$& M7$\,\ldots\,$T0\\ 
$ [4.5]-[5.8]$   & $0.1662$                   &$\Rightarrow$& M8$\,\ldots\,$L6\\
                         &  $0.1662\pm 0.075$& $\Rightarrow$& M0$\,\ldots\,$L7\\ 
$ [5.8]-[8.0]$   & $0.2181$                   &$\Rightarrow$& L0$\,\ldots\,$T4 \\
                         &$0.2181\pm 0.040$  & $\Rightarrow$& M9$\,\ldots\,$T5 \\
$ J -[4.5] $      & $2.1900$                   &$\Rightarrow$ &L3$\,\ldots\,$L5\\
                         & $2.1900\pm 0.175$ & $\Rightarrow$ &M9$\,\ldots\,$T6\\
$ K_{\rm s} - [3.6]$ & $0.8792$                   &$\Rightarrow$& L4 \\
                                  & $0.8792\pm 0.060$ &$\Rightarrow$& L3$\,\ldots\,$L5 \\
$ K_{\rm s} - [4.5]$ & $0.9349$                   &$\Rightarrow$& L4$\,\ldots\,$L5 \\
                                  & $0.9349\pm 0.220$ &$\Rightarrow$& M0$\,\ldots\,$L7 \\
$Y-J_{\rm UKIRT}$    & $1.141\pm 0.3$         &$\Rightarrow$& L \\
$Z-J_{\rm UKIRT}$    & $2.557\pm 0.275$       &\\
$J_{\rm UKIRT}-H_{\rm UKIRT}$    & $0.513\pm 0.4$         &$\Rightarrow$& L \\*[0.4cm]
\multicolumn{3}{l}{\bf T-dwarf test case (4 models):}\\
colour         & \multicolumn{3}{c}{$(m_1 - m_2)_{\rm mean}^{1000K} \,\, \Rightarrow$ \,\,SpT$^{10}$} \\
\hline
$ [3.6]-[4.5]$   & $1.597$                      &$\Rightarrow$& T7.5\\  
                         & $1.597  \pm 0.2900$&$\Rightarrow$& T7$\,\ldots\,$T8\\  
$ [4.5]-[5.8]$   & $-0.339$                     &$\Rightarrow$ & T7\\
                         & $-0.339\pm 0.0800$&$\Rightarrow$ & T7$\,\ldots\,$T7.5 \\
$ [5.8]-[8.0]$   & $0.317$                     &$\Rightarrow$ & L6$\,\ldots\,$T5 \\
                         & $0.317\pm 0.1125$ &$\Rightarrow$ & L6$\,\ldots\,$T6.5 \\
$ J -[4.5] $     & $3.361$                       &$\Rightarrow$  & L9$\,\ldots\,$T8 \\
                        & $3.361\pm 0.2900$&$\Rightarrow$  & L5$\,\ldots\,$T8 \\
$ K_{\rm s} - [3.6]$ & $0.971$                    &$\Rightarrow$  & L4$\,\ldots\,$T4\\
                                  & $0.971\pm 0.2750$&$\Rightarrow$  & L2$\,\ldots\,$T7\\
$ K_{\rm s} - [4.5]$ & $2.568$                   &$\Rightarrow$  & T6.5$\,\ldots\,$T7\\ 
                                 & $2.568\pm 0.2000$&$\Rightarrow$  & T6.5\\ 
$ Y-J_{\rm UKIRT}$   & $0.993\pm 0.175$&$\Rightarrow$  & T\\ 
$ Z-J_{\rm UKIRT}$   & $2.953\pm 0.400$& \\ 
$ J_{\rm UKIRT}-H_{\rm UKIRT}$   & $-0.043\pm 0.75$&$\Rightarrow$  & T\\ 
\end{tabular}
\label{tab:synSpecType}
\end{table}

\subsubsection{General}

We have evaluated our model results in the most conservative way and
despite the differences amongst the simulations in modelling clouds,
they suggest a general agreement in synthetic colours
(Figs.~\ref{fig:ColCol}, ~\ref{fig:ColColZYJ}).  The mean values do
provide a good guidance taking into account the large diversity
regarding the characteristic cloud quantities like the mean particle
sizes and material composition. Of course, uncertainties increase if
the error margins given in Table~\ref{tab:synSpecType} are applied.
The values of the spectral types suggested in
Table~\ref{tab:synSpecType} suffers also the scatter and possible
ambiguity contained in the observed data like e.g. in the IRAC colours
[3.6]-[4.5] or [5.8]-[8.0] as discussed by Patten et al. (2006) which
adds to the model inherent uncertainties.  A better picture appears if
colours are correlated. A comparison with colour-colour plots
(e.g. Lodieu et al. 2007b for UKIRT) shows that the different
simulations reproduce well the spectral classes of the test cases even
in the most uncertain ZYJ colours. Nevertheless, we refrain from the
exercise of back-tracing the T$_{\rm eff}$-values from our synthetic
colours since this would clearly potentiate uncertainties because the
atmosphere simulations used in publications of interest did not treat
the presence of dust at all (in Luhman 1999) or used very simplistic
representations of dust as opacity source (in Golomowski et
al. 2004). 

\section{Discussion}\label{ss:lab}

\subsection{The challenge of phase-transition modelling}

An essential part of modelling clouds in substellar atmospheres is the
description of condensation as a phase-transition gas -
solid/liquid. Two modelling approaches were used in the simulations
compared in this paper: the kinetic approach and the phase-equilibrium
approach.  Condensation occurs in the kinetic approach when a gas
species is supersaturated with respect to its equilibrium concentrate
at given pressure and temperature. Homogeneous nucleation of a
supersaturated species yields macro-molecules, molecular clusters and
eventually nanometer solids. The clusters and nanometer solids could
become seeds for heterogeneous nucleation by surface reactions that
requires lower activation energies than homogeneous nucleation of the
same species directly out of the gas phase. However, the first
condensate will form by homogeneous nucleation. A second condensate
can form at a lower temperature by heterogeneous nucleation on seeds
made of the first condensate or by homogeneous nucleation following
supersaturation of the gas. In the phase-equilibrium approach,
thermodynamic equilibrium is adopted where the Gibbs energy difference
between reagents and products equals zero at fixed values of
temperature, pressure and element composition. A series of these
calculable states when ordered as a function of decreasing temperature
can be viewed as a time sequence of fractional condensation that
predicts the stepped appearance of crystallographically ordered,
chemically stoichiometric solids, i.e.minerals. Equilibrium
condensation models thus predict a sequence of minerals of
systematically different compositions. Yet, the presence of a
particular mineral or mineral assemblage is no proof of equilibrium
condensation.

Laboratory condensation experiments on silicate vapours found that
dissipative structures (i.e. metastable states) appear as highly
disordered, amorphous solids with unique non-stoichiometric
compositions. That is, they have unique Metal-oxide to SiO$_2$ ratios
(M: Mg, Fe, Ca, or Al, and combinations thereof) that match deep
metastable eutectic compositions in equilibrium phase diagrams (Nuth
et al., 1998, 2000; Rietmeijer et al., 1999, 2008). These compositions
are intermediate between equilibrium mineral compositions,
e.g. serpentine dehydroxylate, Mg$_3$Si$_2$O$_7$. With time it would
breaks down to forsterite and enstatite (Mg$_2$SiO$_4$ + MgSiO$_3$)
(Rietmeijer et al., 2002) that both would form at different
temperatures during equilibrium condensation. Equilibrium condensation
is predictable but so again is extreme non-equilibrium condensation of
deep metastable eutectic condensates that are more reactive than
equilibrium minerals. With time, post-condensation thermal annealing
(i.e. ageing) of non-equilibrium condensates will also lead to
thermodynamic equilibrium minerals. Ageing is determined by the
prevailing time-temperature regime (Hallenbeck et al., 2000) and
condensate morphology, i.e. aggregates or dust clumps (Rietmeijer et
al., 1986, 2002). The ageing process will yield minerals at
temperatures below their equilibrium condensation temperature.  Such
processes would require the cloud particles to remain in a certain
thermodynamical state long enough, a situation possibly occurring to
dust trapped in-between convection cells.

Toppani et al.~(2006) demonstrate that their (Mg,Ca,Al,Si)-oxide vapour
condenses to complex hydrates carbonates in a CO$_2$-H$_2$O-rich gas,
and they conclude that this condensation proceeds
near-equilibrium. Their condensation experiments at moderate gas
temperate and low total pressures (1000-1285K, ~0.004 bar) yield many
of the expected equilibrium condensates in crystalline form. However,
both Toppani et al.~(2006) and Rietmeijer et al. (2008) conclude that
the mineralogy of such condensed material can not be understood
without taking into account the influence of kinetics.

\section{Conclusion}
Clouds in the atmospheres of brown dwarfs and gas-giant planets
determine their spectral appearance and influence their 
evolution by altering the atmospheric thermal structure. The challenge of
modelling cloud formation has been approached from very different
perspectives over the past years which leads to the question: Do these
models yield the same results and how much do they differ in predicted
observational quantities?  Five models are compared in this paper to
address these questions. All models emphasise the chemistry of cloud
formation. Considering clouds as the result of a phase transition process (gas to
solid/liquid), the models assuming phase-equilibrium describe the
end-state of the phase-transition process, whereas kinetic models
describe the initial state of the cloud formation process from a
chemical point of view. Which viewpoint is most correct ultimately
depends upon the timescales for the various relevant atmospheric
processes.

The dust cloud models predict generally comparable cloud structures
despite the different approaches, although the results differ
substantially in detail. Opacity relevant quantities like grain size,
amount of dust, dust- and gas-phase composition vary between the
various approaches. Most cloud models agree that small grains composed
mainly of silicates (MgSiO$_2$[s]/Mg$_2$SiO$_4$[s]) populate the upper
cloud layers, whereas iron (Fe[s]) is a major component of the large
grains at the cloud base.  The cloud models agree on the gas-phase
composition in the inner atmosphere only which is too warm for
condensed phases.  All models predict phase-equilibrium here, though
the different models describe the evaporation at different levels of
detail.  Above the cloud, more molecules remain in the gas-phase if
cloud formation is treated in phase-non-equilibrium compared to
results from phase-equilibrium models.  The different results that
arise from differences in cloud modelling are amplified if the entire
atmosphere problem is solved, including radiative and convective
energy transport. The reason is the strong feedback of the clouds on
the $(T, p)$-structure due to the clouds' strong opacity and its high
efficiency in depleting the gas of the atmosphere.

Viewing their spectral appearance, the results of the cloud model
atmosphere codes appear to fall into two categories:\\ -- The {\it
high-altitude cloud models} or {\it extended} (Tsuji$_{\rm
Tcr=1700K}$, Marley, Ackerman \& Lodders, Dehn \& Hauschildt + Helling
\& Woitke) where the dust-to-gas ratio peaks at high altitudes though
at different absolute levels. In these models, small grains ($\langle
a\rangle= 10^{-6}\,\ldots\,10^{-4}\mu$m) are still present well above
the maximum of $\rho_{\rm d}/\rho_{\rm gas}$, hence the gas-phase
absorption is less deep for $\lambda > 1\mu$m.\\ -- The {\it
low-altitude cloud models} or {\it thin} (Tsuji$_{\rm Tcr=1900K}$,
Allard \& Homeier) where the dust-to-gas ratio maximum sits further
inside the atmosphere and no grains populate higher atmospheric layers
above the maximum of $\rho_{\rm d}/\rho_{\rm gas}$. Consequently, the
gas-phase absorption features are much deeper in the mid-IR and IR
part of the spectrum. Consequently, the {\it low-altitude cloud /
thin} model atmospheres appear bluer then the {\it high-altitude cloud
/ extended} model atmospheres.

Comparing synthetic photometric fluxes and colours from different
model atmosphere codes illustrates the current range of uncertainty,
or error bar, for theoretical predictions. These error bars are worst
cases. They are derived from a group of different models and not from
only one particular family of models.  In the most conservative case,
the maximum differences in photometric fluxes, $\Delta_{\rm max}[\log
F_{\lambda}]$, amongst the models are between 1\% and
30\%. $\Delta_{\rm max}[\log F_{\lambda}]$ increases to 50\% for the
T-dwarf test case in the WFCAM UKIRT wavelength intervals. This
translates into an uncertainty in apparent magnitudes for the L-dwarf
test case of $0.25 < \Delta m < 0.875$. $\Delta m$ increases to 1.375
for the T-dwarf test case in the WFCAM UKIRT filter system.

We conclude that every comparison with observations should ideally
involve models from different groups.  This would allow the
determination of a synthetic error bar in determinations of
fundamental quantities like T$_{\rm eff}$, $\log g$, and metallicity.
Ultimately comparison of models to objects with independently
constrained properties (from orbital motion and bolometric luminosity,
for example) will elucidate the modelling approaches that most
accurately capture the relevant physics.  Other possibilities for
tests include objects in young stellar clusters which have well
constrained ages and metallicities.

Future works on cloud formation need to seek more support in
laboratory astrophysics. Hydrodynamic modelling of ultra-cool
atmospheres will provide the opportunity to study the dynamic
processes of cloud formation. They need to including a
consistent description of the chemical formation processes and simultaneously 
address the challenge of a turbulent fluid field.

\section*{Acknowledgements}
We thank the anonymous referee for the valuable report.  ChH thanks
Alkes Scholz and S\"oren Witte for helpful discussions on the
manuscript.  We thank the participants of the workshop {\sf From Brown
Dwarfs to Planets: Chemistry and Cloud formation} which was supported
by the Lorentz Center of the University Leiden, Nederlandse
Organisatie voor Wetenschappelijk Onderzoek, The Netherlands research
School for Astronomy, the Scottish University Physics Alliance, and
European Space Agency.

\end{document}